

\magnification=1200
\baselineskip=16pt

\pageno=1
\hskip 10cm
ICN-UNAM 9416
\vskip1pc

\centerline{\bf THE CONSTRAINTS IN SPHERICALLY SYMMETRIC}
\centerline{\bf  CLASSICAL GENERAL RELATIVITY  II}
\vskip2pc
\centerline{\bf IDENTIFYING THE CONFIGURATION SPACE:}
\centerline{\bf A MOMENT OF TIME SYMMETRY}
\vskip2pc
\centerline {\bf Jemal Guven$^{(1)}$ and Niall \'O Murchadha$^{(2)}$ }
\vskip1pc
\it
\centerline {$^{(1)}$Instituto de Ciencias Nucleares}
\centerline {Universidad Nacional Aut\'onoma de M\'exico}
\centerline {A. Postal 70-543. 04510 M\'exico, D. F., MEXICO} \rm
\centerline{(guven@roxanne.nuclecu.unam.mx)}
\vskip2pc
\it
\centerline {$^{(2)}$ Physics Department}
\centerline {University College Cork}
\centerline {Cork, IRELAND} \rm
\centerline{(niall@iruccvax.ucc.ie)}
\vfill\eject
\centerline{\bf Abstract}
\vskip1pc
{\leftskip=1.5cm\rightskip=1.5cm\smallskip\noindent
We continue our investigation of the configuration space
of general relativity begun in I (J. Guven and N. \'O Murchadha,
gr-qc/9411009). Here we examine
the Hamiltonian constraint when the spatial geometry is
momentarily static (MS). We begin with a heuristic description of
the presence of apparent horizons and singularities.
A peculiarity of MS configurations is that
not only do they satisfy the positive quasi-local mass
(QLM) theorem, they also satisfy its converse: the QLM is
positive everywhere, if and only if the (non-trivial)
spatial geometry is non-singular. We derive an analytical
expression for the spatial metric in the neighborhood of a
generic singularity. The corresponding curvature singularity
shows up in the traceless component of the Ricci tensor.
As a consequence of the converse, if the energy density of matter is
monotonically decreasing, the geometry cannot be singular.
A supermetric on the configuration space
which distinguishes between singular geometries
and non-singular ones is constructed explicitly.
Global necessary and sufficient criteria
for the formation of trapped surfaces and singularities
are framed in terms of inequalities which relate some
appropriate measure of the material energy content
on a given support to a measure of its volume.
The sufficiency criteria are cast in the form: if the material energy
exceeds some universal constant times the proper radius, $\ell_0$, of the
distribution, the geometry will possess an apparent horizon for
one constant and a singularity for some other larger constant.
A more appropriate measure of the material energy for
casting the necessary criteria is the maximum value of the
energy density of matter, $\rho_{\rm Max}$: if
$\rho_{\rm max}\ell_0^2 <$ some constant the distribution of matter
does will not possess a singularity for one constant and
an apparent horizon for some other smaller constant.
These inequalities provide an approximate characterization of
the singular/non-singular and trapped/non-trapped
partitions on the configuration space. Their strength
is gauged by exploiting the
exactly solvable piece-wise constant density star as a template.
Finally, we provide a more transparent derivation
of the lower bound on the binding energy conjectured by
Arnowitt, Deser and Misner and proven by Bizon, Malec and \'O Murchadha
and speculate on possible improvements.

\smallskip}
\vfill
\eject

\centerline{\bf 1. INTRODUCTION}
\vskip1pc

This is the second paper in a series in which we examine the
solution of the constraints in general relativity under the
restriction that both the spatial geometry and the extrinsic curvature be
spherically symmetric. In the first paper (hereafter referred
to as paper I), we examined various universal features
of the constraints [1]. In particular, it was shown following [2]
how the constraints can be recast in a symmetrical form
with respect to the optical scalars of the theory. We introduced
an appropriate large class of foliations which are in some sense
the natural foliations of a spherically symmetrical spacetime.
We discussed the positivity of a quasi-local mass and related local
bounds on the canonical variables of the theory. Finally,
a tentative realization of the configuration space was proposed.

In this paper we will specialize to solutions of the
constraints when the geometry is momentarily static (MS)
and asymptotically flat.\footnote *
{A static solution is a solution possessing a surface forming timelike
Killing vector. It is easy to show that the spacelike surface orthogonal
to the Killing vector satisfies $K_{ab}=0$.}
The extrinsic curvature and the radial material current now both vanish,
$K_{ab}=0= J$
and the radial momentum constraint becomes vacuous.
We follow the notation established in paper I.
The proper (or geodesic) radius $\ell$ is treated as
the independent radial coordinate in our parametrization of the spatial
geometry. The Hamiltonian constraint then provides an ODE for the
circumferencial radius as a function of $\ell$. We defer the examination
of non-vanishing extrinsic curvature to a third paper (III) [3].

Whether the solution remains static, subject to the initial
conditions $K_{ab}=0$ so that $\dot K_{ab}=0$ will depend on
the stresses which only show up in the dynamical equations. What
is true is that if one is to construct
globally static solutions, they had better satisfy the
Hamiltonian constraint with $K_{ab}=0$ to begin with.

While the constraints simplify enormously when $K_{ab}=0$,
many if not all of the subtleties involved in the solution of the
initial value problem obligingly manifest themselves.
This makes such configurations important as a theoretical proving
ground.

Another feature of these solutions which makes them
interesting is that they are the only geometries which
simultaneously satisfy the Lorentzian and Euclidean constraints.
In the semi-classical approximation to tunnelling,
they correspond to the classical turning points. They therefore
represent the boundaries between the classically
allowed and forbidden regimes [4].

Our goal will be to provide a characterization of the
configuration space, ${\cal C}$, of all spatial three-geometries
satisfying the Hamiltonian constraint essentially by whatever means
we can. From one point of view, of course, ${\cal C}$ could be
characterized by whatever material degrees of freedom we introduce and,
as such, is trivial even if the corresponding Hamiltonian
is not [5]. If matter is modelled by a scalar field, for example,
${\cal C}$ will simply be all spherically symmetric scalar
fields  with an appropriate falloff.
What is ignored here, however, is that to
each point in the configuration space there also corresponds
a spatial geometry. In particular, some configurations will
correspond to geometries possessing apparent horizons, with a
proper subset of these corresponding to geometries possessing singularities.
The boundaries in ${\cal C}$ partitioning
geometries with apparent horizons or singularities
from those which do not are important landmarks on this
space.

In Sect.2 we discuss generic analytical properties of the
Hamiltonian  constraint. We choose, as our independent
data, a spherically symmetric energy density
distribution. This consists of a non-negative function $\rho$, given as a
function of the proper radius $\ell$ on the positive real axis,
$[0,\infty)$. The Hamiltonian constraint is then solved
to give the spatial geometry. This is to be regarded as a kinematical
quantity, entirely determined by the source distribution. We are not fussy
about the differentiability of $\rho$, we would be happy with $C^N$ where
$N$ is any integer in the interval $[0, \infty]$.

We suppose that the asymptotically flat geometry is regular at its
center. Regular solutions of the Hamiltonian constraint
are determined uniquely once $\rho$
is specified. For some choices of $\rho$, however,
we find that the circumferential radius $R$ goes to zero at a finite value
of $\ell = \ell_S$.

Let us suppose we scale some small $\rho$, say
$\rho_0$. Initially, the circumferential radius $R$ increases
monotonically with the proper radius. As the scale
is raised a critical point is always reached in which the
geometry develops an apparent horizon. When $K_{ab}=0$, this
corresponds to the development of an extremal embedded two-sphere.
At this critical scaling, the horizon will be degenerate.
As $\rho$ is scaled further, the horizon will
bifurcate into an outer minimal two-sphere and an inner maximal two-sphere.
The interpolating two-spheres which foliate the spatial region between
the two will all be trapped. Typically, if $\rho_0$ is scaled further,
the geometry will become singular at some other critical scaling
corresponding to the pinching off of the minimum two-sphere to $R=0$
at $\ell = \ell_S$.
All points on this two-sphere would then be identified,
the geometry would possess a bag of gold at its center.
This is the only way that the spatial geometry could possess a
singularity. These three-dimensional singularities are the only
ones we discuss. These form only a subset of the possible four-dimensional
singularities that could arise when regular initial data is evolved using
the dynamical equations to find a solution of the Einstein equations. One
particular class of singularities (which are simultaneously three- and
four-dimensional) we do not discuss are those which arise when $\rho$ at
a point becomes unboundedly large.

Physically, it might be argued that the singular geometries we consider
are irrelevant from the point of view of the asymptotically flat
spacetime as they are disconnected from it. It could also be
argued, however, that this feature is an artifact of the
spherically symmetric model. If the symmetry is relaxed, the
singular structures which arise will, typically, not disconnect the
spatial geometry. This is not, however, our justification for
considering singular structures. They are important for the reason
indicated by the scaling argument. An infinesimal change in
$\rho$ can convert a regular geometry into a singular one. The
description of the configuration space is incomplete if
such singular geometries are ignored.

A remarkable feature of spherically symmetrical geometries is
the positivity of a well defined
quasi-local mass (QLM) whenever the spatial geometry
is regular and the sources satisfy the dominant energy
condition [1,6]. Its converse is not, however, generally true ---
there are singular solutions of the constraints with a
positive everywhere quasilocal mass [3].
A peculiarity of MS configurations is that
they do satisfy the converse of the positive QLM theorem. If the QLM is
positive everywhere, the geometry is necessarily non-singular.
We can now exploit the definition of the QLM to derive an analytical
expression for the local form of the spatial metric
in the neighborhood of a generic singularity. Surprisingly, the
QLM remains finite at the singularity.
The constraints guarantee that the scalar curvature is also
finite everywhere, even if the geometry is singular. If there is a
curvature singularity, it must show up as a
divergence in the traceless component of the Ricci tensor.
Non-generic radial density profiles exist in which
the geometry degenerates without any
curvature singularity where $R$ pinches off. The interior region
behind the singularity can be interpreted as a regular
closed universe.

Our understanding of the analytic form of the metric in the neighborhood
of a singularity, facilitates the identification of
a supermetric on ${\cal C}$ in the manner  of DeWitt [7].
Specifically, in Sect.3 we identify a non-ultralocal metric which
assigns  a finite norm to regular geometries and an infinite
norm to singular geometries. With respect to this metric, the
boundary  in ${\cal C}$ separating singular geometries from
non-singular ones can be consigned to infinity.

A surprising consequence of the converse of the
positive QLM theorem is that if $\rho$ is strictly monotonically
decreasing, the geometry cannot be singular (if it is regular at the origin).
What this demonstrates is that $\rho$ can be scaled by
an arbitrarily large factor on a given support
and yet the geometry remain regular everywhere.
This counter-example demonstrates clearly
that the scaling argument breaks down as a description of the
development of singularities. For while an apparent horizon always
develops at some critical scaling, a singularity does not necessarily
develop. Such geometries may not be of great physical interest because
we would expect that any classical singularity that might
form in such a collapsing configuration would form at its center
where we have assumed that the geometry is regular.
However, monotonicity does provide a useful boundary in
${\cal C}$ separating geometries with very different
properties with regard to singularity formation.

We will illustrate these points explicitly in Sects.4 and 5 using
a piece-wise constant density star.

We discuss the constant density star in greatest detail.
The model does not admit curvature singularities;
its only pathologies are metric degeneracies. Because of this
one needs to be extra wary of making generalizations.
We illustrate how a mechanical analogue can be
used to study the solution space.

The local measures of energy and size
which are most appropriate for measurements performed outside the star
consist respectively of the QLM $m$ and $R$ at the surface of the star.
What is clear already in this model, however, is that
$m$ provides a very poor measure
of the material energy content just as $R$ becomes a poor
measure of the size when the geometry is highly curved.
In particular,  $m$ (or $R$ ) may be infinitesimally small in
geometries possessing apparent horizons.

It is true that an observer living outside the star
does not have any choice but to exploit these `local'
measures. Our goal, however, is to
chart the configuration space, above all, in the
neighborhood of configurations with
apparent horizons and singularities.
Because we are not asking the kind of question a remote
orbiting observer in one of these geometries might ask, there is no
reason why we should handicap ourselves with variables
more appropriate to another scenario.
The geometry behind a horizon is not out of bounds.
In is in this spirit that we identify
our global measures of energy content and size.

A measure which does appear to correspond
quantitatively  to the material energy content is the
integrated energy density over the spatial volume, $M$.
The most appropriate measure of size is the proper radius
of the support of matter, $\ell_0$.
The monotonicity of these two variables means that they fare better than
$m$ and $R$ in the strong field region we are
interested in. We examine in detail
the dependence of $M$ on $\ell_0$.

A two-density model is discussed in Sect.5. We demonstrate
explicitly that when the inner density exceeds the outer one, the model
can never possess a singularity no matter how great the density or
large its support. If, on the other hand, the outer density is the greater,
singularities can appear if the star is large enough. If the
outer density is tuned appropriately, the curvature
singularity becomes a metric degeneracy.

In Sect. 6 we follow Bizon, Malec and \'O Murchadha (BM\'OM) by
concentrating on the formulation of global
necessary and sufficient conditions determining when
the spatial geometry will possess apparent horizons and singularities
[8,9,10]. The motivation behind this work was Thorne's
hoop conjecture [11].
These conditions are cast as inequalities involving
appropriate global measures of the energy content of the matter distribution
and of its support.

The sufficient conditions can be cast in the form:
if $M\ge  C\,\ell_0$, where $C$ is a constant, then the star
possesses a trapped surface for one constant, and a singularity for some
larger constant within the sphere of proper radius $\ell_0$.
On the subset which corresponds to a
monotonicically decreasing density these inequalities
can be strengthened. However, the
latter inequality now provides a universal bound on $M$.
To understand how this comes about suppose we are given some
monotonically decreasing density, $\rho_0$. While one might imagine that
by scaling $\rho_0$ by some large constant we
could make $M$ arbitrarily large, this is misleading because the
volume element appearing in $M$ contains $R^2$. Given $\rho_0$,
the constraint determines the value of $R$. When we solve the
constraint with the scaled value of $\rho$, the net effect is to
decrease $R$. This can occur in such a way that $M$, in fact,
saturates.

In the original work of BM\'OM, conformally flat coordinates
were exploited. The attraction of these coordinates is that
they do not rely on the spherical symmetry of the problem. The
super-Hamiltonian constraint reduces to a non-linear elliptic
PDE on flat $R^3$. This is offset, however,
by the fact that the background geometry is no longer the
physical geometry and one must contend with
unphysical bifurcations in the solution space
of the elliptic equation. What is more, the existence of the
inequalities discovered by BM\'OM is not obvious when the constraints are
expressed in this way. Their derivation demonstrated  a considerable level
of ingenuity. By expressing the constraints directly in terms of
physical quantities, we are able to provide
very simple alternative rederivations of the BM\'OM inequalities
which are both more economical and transparent than theirs.

The shortcomings of $M$ become apparent
(even in the constant density model) when we
attempt to provide necessary conditions on the
specification of the boundaries partitioning the configuration space.
For, if $M\le C\, \ell_0$, $C$ a constant, on the surface of the star,
for some constant no matter how small, even if it is true that the
surface is neither trapped nor singular, this is not true of the
interior. What is worse, we find it impossible to provide even this weak
form of necessary condition involving the surface geometry
when currents flow [3]. This contrasts dramatically with
the relatively straightforward generalization of the
sufficiency conditions in this context.

A more appropriate measure of the energy content of the star
for phrasing the necessary condition is provided by the
maximum value of the energy density, $\rho_{\rm Max}$. In Sect.7, we
demonstrate that if $\rho_{\rm Max}\ell_0^2 < {\rm constant}$,
then the interior is free of trapped surfaces for one constant, and free
of singularities for some other larger constant. The
proof requires the introduction of a battery
of simple Sobolev and other inequalities.
These bounds possess extremely  non-trivial generalizations to $J\ne 0$.

As a demonstration of the effectiveness of our method, in Sect.7
we rederive the BM\'OM proof of Arnowitt deser and Misner's
conjectured inequality placing a lower bound on the binding
energy of a compact spherically symmetric system [10].

\vfill
\eject

\noindent{\bf 2 GENERIC ANALYTIC FEATURES OF SOLUTIONS}
\vskip1pc

The Hamiltonian constraint provides a second order non-linear
ODE for $R$ which is regular everywhere except at $R=0$\footnote *
{The Hamiltonian constraint will also assume this form in either
of the gauges, $K_R=0$ or $K_{\cal L}+0.5 K_R=0$ which mimic
a MSC (see paper I).
In the latter case, however, apparent horizons will  not coincide
with minimal surfaces and we need to be careful when we carry
over the MSC analysis.}

$$R R^{\prime\prime}-{1\over 2}(1-(R^\prime)^2)+
4\pi  R^2 \rho=0\,.\eqno(2.1)$$
The closure of the geometry at its base point
$R(0)=0$, is the only boundary condition we are free to impose.
We will require that the geometry be regular at this point[1], so that

$$R^\prime(0)=\pm 1\,.\eqno(2.2)$$
Otherwise, it is clear that $R^{\prime\prime}$ would be infinite
and the geometry singular at the origin.
$R$ can be either positive or negative --- the
physical metric only depends on $R^2$.
However, the only way $R$ can change sign
is by passing though a singularity or
degeneracy with $R=0$ at some non-vanishing
value of $\ell$. If the geometry is regular, the sign of $R$ will
not change. Our convention will be to choose the positive sign
at $\ell=0$.

Because Eq.(2.1) is singular at the origin, we need
to differentiate the equation $n-1$ (not $n-2$)
times to determine the $n^{\rm th}$
derivative at this point, $R^{(n)}(0)$.  When we do this,
we obtain the power series expansion for $R(\ell)$
in the neighbourhood of $\ell=0$,

$$R(\ell)=\ell - {4\pi  \rho(0)\over 9}\ell^3 +{\cal O}(\ell^4)\,.\eqno(2.4)$$

The most convenient way to solve Eq.(2.1) is to exploit the
fact that the equation possesses a first integral
equating the quasi-local mass defined at a given value of $\ell$ in terms
of the interior energy content. In paper I, this was done
in a completely general gauge invariant
context in which neither $K_{ab}$ nor $J$ need be zero [1].
If we set both $K_{ab}$ and $J$ equal to zero in Eqs.(4.5) and (4.6) of [1]
we obtain\footnote {$^\dagger$}
{That this is also true in the polar gauge
discussed in paper I is obvious.
It is also true in the other mimicking gauge ($\alpha=0.5$)
but we must remember that what we call $m(\ell)$ in Eq.(2.5) is no longer
the QLM.}

$$(R^\prime)^2=1- {2m(\ell)\over R}\,,\eqno(2.5)$$
where $m(\ell )$ is given by
$$m(\ell)=4\pi\int_0^\ell d\ell \,R^2 R^\prime \rho\,.\eqno(2.6)$$
Eqs.(2.5) and (2.6) are a first integral of Eq.(2.1) which implement the
condition of  regularity at the origin, Eq.(2.2), explicitly.

Outside the support of matter $m(\ell)$ coincides with the ADM mass,
$m_\infty$ and

$$(R^\prime)^2=1- {2m_\infty\over R}\,,\eqno(2.5^\prime)$$
If $\rho=0$ everywhere, $m=0$ and space will be
flat, $R(\ell)=\ell$.
If $m_\infty$ is positive and $R$ is large
the second term on the right hand side of ($2.5^\prime$)
becomes negligible and $R\sim \ell $ to leading order.
Asymptotic flatness is guaranteed by the constraint.
The `boundary condition' at infinity
does not need to be prescribed by hand.
The ADM mass is encoded in the next to leading order,

$$R\sim \ell -m_\infty \ln \ell\,.\eqno(2.7)$$
Because $m_\infty$ is positive $R^\prime$ must tend to one from below.

The power series (2.4) would appear to suggest
that, once regularity at the origin is implemented, the
solution will remain regular (regardless of the singularity of the
differential equation at $R=0$) so long as $\rho(\ell)$
remains finite. Certainly, if $\rho$ and its support are small,
on physical grounds we would expect the solution to be regular
with $R$ increasing monotonically with $\ell$ between $\ell=0$ and
infinity. If $\rho$ (or its support) is sufficiently large, however,
$R(\ell)$ need not continue to increase monotonically with $\ell$.
We can imagine doing this by scaling on some initial small value.
Beyond some critical scale factor,
the curvature of the geometry will be sufficiently large that $R^\prime$
will vanish at some value of $\ell$, an apparent horizon forms,
outside which $R(\ell)$ begins to decrease. Note that this behavior
is consistent with the convexity of $R(\ell)$ given by (2.4) in the
neighborhood of the origin. We will be more precise below.

What can go wrong beyond this point is that $R$ might not recover
but will continue decreasing until it reaches zero (where the
constraint equation is singular) at some non-vanishing value of $\ell$. If
we extrapolate the first two terms in Eq.(2.4)
to larger $\ell$, we see that
$R(\ell)$ returns to zero when $\rho(0)\ell_0^2= 9/4\pi$ ---
in reasonable agreement with the value for the constant density star
discussed in Sect.4. The important point is that this is the only way that
the spatial geometry can be singular.\footnote *
{Even when $\rho$ is large, $R(\ell)$
cannot diverge at any finite value of $\ell$. This is because
this would require $R^\prime$ to diverge whereas the only place
it can diverge is at the singular point of the
differential equation $R=0$. }

In Sect.5 of paper I, we proved the positivity of the QLM, $m$: if
the geometry is regular and the weak energy condition is
satisfied everywhere then $R^{\prime2}\le 1$ everywhere
and, as a consequence, $m\ge 0$.\footnote {$^\dagger$}
{The contrapositive of this statement is that
if $m(\ell)<0$ at some $\ell$, the geometry must become singular at some
point beyond $\ell$.}

Obviously, $m(\ell)$ goes to zero as one approaches the origin.
It does so faster than $R$ so that $R'$ approaches unity. Starting from
the origin and moving outwards, we see that $m(\ell) = 0$ and $R' = 1$
until we run into some nonzero $\rho$. If the support of $\rho$ is
bounded away from the origin we get a locally flat region around the
origin. Once we run into nontrivial $\rho$  we have  that $m(\ell)$
must increase to reach some positive value. If $m(\ell)$ is bounded from
below by some  positive constant value for all positive values of $\ell$
outside a neighbourhood of the origin then $R(\ell)$ cannot become small in
this region if $K_{ab}=0$. This is because this would imply that the second
term on the right hand side of Eq.(2.5) must exceed the first at some point
which is impossible. Thus the geometry cannot become singular.

We can therefore claim that $m(\ell)>0$ everywhere except at the origin
or in a neighbourhood of it if and only if the geometry is
non-singular. This neighbourhood, of course, covers the whole space only in
the trivial case where $\rho \equiv 0$ and we have flat space. The set of
geometries with $m(\ell)=0$ somewhere away from the origin is a set of zero
measure and will be treated separately below after we have discussed the
constant density star.

What is the analytical description of a generic singularity?
Suppose that  $R^{\prime}(\ell) < -1$, or
equivalently $m(\ell)<0$ at some point $\ell$. Eq.(2.1) tells us that
$R^{\prime\prime} < 0$. Hence $R^{\prime}$ must become more negative.  $R$ must
return to zero at some finite value $\ell_S$ beyond  $\ell$,
$R(\ell_S)=0$. If $\rho$ is bounded on the interval
$[0, \ell_S]$ it is clear that $m(\ell_S)$ reaches a finite negative value.
Let us assume that $R$ goes to zero like $(\ell_S-\ell)^{\gamma}$, where
$\gamma$ is some positive constant. We then expect $R'$ to behave like
$(\ell_S-\ell)^{\gamma - 1}$. From Eq.(2.5) we see that  $m(\ell)$ goes like
$(\ell_S-\ell)^{3\gamma - 2}$ and that $m'(\ell)$ should go like
$(\ell_S-\ell)^{3\gamma - 3}$. However, if we differentiate Eq.(2.6) we find
that  $m'(\ell)$ equals $4 \pi R^2 R' \rho$ and this means it should behave
like $(\ell_S-\ell)^{3\gamma - 1}$. This is inconsistent. The only way out
is if $3\gamma = 2$. This means that $m(\ell)$ goes to a constant and when
we differentiate we do not get the  $3\gamma - 3$ term.

 Near
$\ell_S$, the second term  in Eq.(2.5) will dominate so that

$$(R^\prime)^2\sim -{2m(\ell_S)\over R}\,.\eqno(2.8)$$
This equation can be integrated to give

$$R\sim \left({3\over 2}\right)^{2/3} \left(-2m(\ell_S)\right)^{1/3}
(\ell_S-\ell)^{2/3}\,,\eqno(2.9)$$
entirely consistent with the argument above. $R^\prime$ becomes unbounded at
$\ell_S$ as $-(\ell_S-\ell)^{-1/3}$.
If, therefore, $R^{\prime2}>1$ ($R^\prime <-1$)
anywhere it must subsequently diverge.
It is clear, from (2.6) that not only
does $m$ remains finite at $\ell_S$, so also does the integrand
$\rho R^2 R^\prime$.

All points with $\ell=\ell_S$ are zero distance apart so they must be
identified.  The geometry now contains one of Wheeler's  bags of
gold [12] behind the singularity. Because the singularity cuts off
the interior geometry, it is customary in the present context to discard
the solution as unphysical. The exterior does, however, also
contain this singularity.

Generically $R^\prime$ will diverge as we approach a `singularity'.
We will call this a strong singularity. If, however, we fine tune
the density so that $m(\ell_S)=0$ then $R^\prime(\ell_S)=-1$,
and $R^{\prime\prime}(\ell_0)=0$. The bag closes smoothly.
and there are no singularities in the curvature.
We can always approximate such a geometry arbitrarily
closely by a regular geometry. We will refer to this as a weak singularity.

A beautiful consequence of the converse of the positive QLM
theorem is that it
permits us to identify an important class of non-singular geometries.
We note that, by a simple integration by parts, $m(\ell)$ can
be expressed in the alternative form

$$m(\ell)={4\pi\over3} R^3\rho -
4\pi\int_0^{\ell} d\ell\, R^3 \rho^\prime\,.\eqno(2.6^\prime)$$
If the energy density is monotonically decreasing, $\rho^\prime\le0$, then

$$m(\ell)\ge {4\pi\over3} R^3\rho \,.\eqno(2.10)$$
In particular, if $\rho$ is a monotonically decreasing function of
$\ell$, and it is strictly monotonically decreasing in some neighbourhood
of the origin then $m(\ell)>0$, and so the geometry is necessarily
non-singular.

We note that neither the converse of the positivity of the QLM nor
the consequence survive when $J\ne 0$.
In general, a positive value of $m(\ell)$
everywhere does not guarantee the non-singularity of the geometry.

What is the geometrical nature of a generic spatial
singularity? Even though the geometry might be singular, we assume
$\rho$ is always finite. The constraint therefore
guarantees that the scalar curvature remains finite. The metric
singularity must therefore show up in the Ricci tensor. We can express

$${\cal R}_{ab}= {\cal R}_{\cal L} n_a n_b + {\cal R}_R (g_{ab}-n_an_b)\,,
\eqno(2.11)$$
in the same way as we did $K_{ab}$. All curvature scalars can be expressed
in terms of ${\cal R}_{\cal L}$ and ${\cal R}_R$.
In appendix I, we show that

$$\eqalign{{\cal R}_{\cal L}
		 =&{{\cal R}\over 2} -{1\over R^2} (1- R^{\prime2})\cr
{\cal R}_R=&{{\cal R}\over 4} + {1\over 2 R^2} (1- R^{\prime2})\,.\cr}
\eqno(2.12a,b)$$
While the scalars $R_{\cal L}$ and $R_R$ both diverge as one
approaches a strong singularity, the sum ${\cal R}=
{\cal R}_{\cal L}+ 2{\cal R}_R$ remains finite. In fact, as we
approach a strong singularity,
${\cal R}_{{\cal L}, R}\sim \pm R^{\prime2}/R^2 \sim
\pm (\ell_S-\ell)^{-2}$.
The singularity need not occur on the support
of matter. At a weak singularity ${\cal R}_{ab}$ remains finite.

\vskip2pc
\noindent{\bf 3 TO THE SINGULAR BOUNDARY OF CONFIGURATION SPACE}
\vskip1pc

\noindent{\bf 3.1 Supermetrics}
\vskip1pc

In this section we construct a supermetric on the configuration space,
${\cal C}$, of spherically symmetric asymptotically flat geometries
satisfying the Hamiltonian constraint at a moment of
time symmetry. Such geometries are completely characterized by the function,
$\rho(\ell)$. We can introduce a flat line element on ${\cal C}$

$$||\delta \rho||^2 =\int_0^\infty d\ell\, \delta\rho(\ell)^2\,.\eqno(3.1)$$
which is ultra-local with respect to $\rho$. The
flatness of the metric makes life very convenient in that
it is then simple to determine the `distance' between
finitely separated geometries. The shortcoming of the metric, (3.1),  is
that it does not tell us anything about the structure of the
underlying geometry.

To do this, we need to construct a line element which depends
explicitly on the geometry. In the full theory, the commonly accepted
choice is the ultra-local DeWitt supermetric

$$||\delta g ||_{DeW\,0}^2=
\int d^3 x\sqrt{g} G^{ab\,cd} \delta g_{ab} \delta g_{cd}\,,
\eqno(3.2)$$
where $G^{ab\,cd}={1\over2}\left( g^{ac} g^{bd} + g^{ad} g^{bc}\right)
+ C g^{ab} g^{cd}$
and $C$ is some constant which is determined by the
requirement that $G^{ab\,cd}$ be positive definite.
On the subset of spherically symmetric geometries, $||\delta R ||_{DeW\,0}$
is given, up to an irrelevant constant, by

$$||\delta R ||_{DeW\,0}^2
=\int_0^\infty d\ell\, \delta R(\ell)^2\,,\eqno(3.3)$$
which, remarkably, is also flat. Note that the DeWitt
supermetric is not itself flat. The metric it induces on
the subspace of spherically symmetric geometries is flat.

We note that with respect to any other parametrization of the spatial
metric, the form assumed by the DeWitt supermetric
will be more complicated. For example, with respect
to the Schwarzschild parametrization (when valid) in the notation of I:

$$||\delta {\cal L} ||_{DeW\,0}^2
=\int_0^\infty dR\, {\delta {\cal L}(R)^2\over {\cal L}}\,.\eqno(3.4)$$
Unfortunately, no matter how we parameterize it, the norm on any
asymptotically geometry, even the flat geometry, is infinite with respect to
the ultra-local metric Eq.(3.2) simply because $R\sim\ell$ at infinity.
In fact the distance between any two geometries with
differing values of $m$ will also be infinite with respect to
$||\delta R ||_{DeW}$. The reason for this is that the
asymptotic behavior given by Eq.(2.7) implies that
$R_1-R_2 \sim (m_{\infty1}-m_{\infty2}) \ln \ell$ which is not
square integrable. It is not, however, clear if this is
anything to worry about. $m_\infty$, after all, is conserved and,
classically at least, different values of $m$ correspond to subsets
of ${\cal C}$ which do not intersect. Even
quantum mechanically, it would appear that the
theory must be constructed with a fixed value of $m_\infty$.

It is, however,  simple to construct a metric
which gives a finite norm between geometries with
different $m_\infty$'s. We do this by relaxing the
criterion of non-ultralocality  by considering a metric
involving a finite number of derivatives of $R$.
If this seems like a high price to pay, one should note that
in terms of $\rho$, even the DeWitt ultralocal metric is
extremely non-local. Consider the metric

$$||\delta R ||_{DeW\,1}^2 =\int_0^\infty d\ell |\delta R^\prime(\ell)|^2\,.
\eqno(3.5)$$
Now $R_1^\prime-R_2^\prime \sim (m_{\infty1}-m_{\infty2}
)/\ell$	which is square integrable. Even still, the norm of any geometry is
still infinite with respect to the non ultra-local metric we have introduced.
Furthermore, none of the metrics we have considered so far,
discriminates against the kind of singular geometries
which show up as solutions of the constraints.
In particular, it is clear from Eq.(2.9) that
$R'$ is square integrable over any
compact interval, whether the geometry is singular or not.
The simplest metric that solves both problems is one
which is second order in derivatives \footnote *
{In general, on any compact support, these metrics satisfy the
sequence of (Poincar\'e) inequalities:
$$||\delta R ||_{DeW\,0}\le ||\delta R ||_{DeW\,1}\le
||\delta R ||_{DeW\,2}\le \cdots \,.$$}

$$||\delta R ||_{DeW\,2}^2 =\int_0^\infty d\ell
|\delta R^{\prime\prime}
(\ell)|^2\,.\eqno(3.6)$$
Now  $||\delta R ||_{DeW\,2}$ is finite on any
non-singular geometry that satisfies the constraints.
Eq.(2.9) indicates that, with respect to this metric,
not only are strongly singular geometries pushed out to
infinity, they are also rendered infinitely distant
from any non-singular geometry.
What is also very attractive is that, with respect to this metric, the
norm of the flat geometry $R(\ell)=\ell$ is zero. However, this
might be disconcerting for those who advocate
a zero geometry as the ground state. While the flat metric in
$\rho$ also does this, it fails to penalize singularities.

It is possible to introduce a metric which is geometrically
more satisfying, which is also second order in derivatives by casting
it directly in terms of differences in the
curvature scalars. Unfortunately, the price
we pay is that the resulting metric is no longer flat.
In addition, the simplest such choice,

$$||\delta {\cal R}||^2=\int_0^\infty d\ell R^2
|\delta {\cal R} (\ell)|^2\,,\eqno(3.7)$$
is a step back because it fails to discriminate against
singular geometries. This is solved by considering the
metric ($C=0$ in the DeWitt supermetric)

$$\eqalign{||\delta {\rm Ricc}||^2= &\int_0^\infty
d\ell R^2 G^{ab\,cd}\delta {\cal R}_{ab}\delta {\cal R}_{cd}\cr
=&\int_0^\infty d\ell R^2
\Big[|\delta {\cal R}_{\cal L} (\ell)|^2
+2|\delta {\cal R}_R(\ell)|^2\Big]\,.\cr}\eqno(3.8)$$

\vskip1pc
\noindent{\bf 3.2 The Singular Boundary}
\vskip1pc

Perhaps, the most important partition of ${\cal C}$ is into
geometries which possess one or more singularities and those which do not,
${\cal C}_0$. ${\cal C}_0$ is clearly open.
Within ${\cal C}_0$, lie all geometries corresponding to (strictly)
monotonically decreasing $\rho(\ell)$. With respect to any of the
metrics we have introduced on ${\cal C}$,
two stars with monotonically decreasing energy density profiles
can be arbitrarily far apart. ${\cal C}_0$ is therefore unbounded,
at least in some directions, with respect to this metric.
The boundary ${\cal S}$ appears then to be
disconnected. Our task will be to begin to characterize it.

It is not clear if we should reject singular geometries as acceptable
points on the configuration space. Should we treat the
surface separating singular geometries from non-singular ones as the
boundary of the physical configuration space?
If this is so then we are faced with a major problem. Technically,
it appears unlikely that we will ever be in a position to
identify ${\cal S}$ exactly. In the quantum theory,
we would be required to implement a
boundary condition on the wavefunction on ${\cal S}$.
What is to be hoped for is that the potential determining
the dynamics in the configuration space will be
sufficiently steep in almost all directions
as we enter the grey area in the neighborhood of
this boundary that wave-packets will get
reflected back onto the support of `good' geometries.
An analogy which might be useful is the manner in which
all wave packets which enter the Taub channels in
homogeneous mini-superspace quantum gravity appear to get
reflected back into the generic mixmaster mini-superspace [13].

There is another natural partition within
${\cal C}_0$, with boundary ${\cal T}$ into geometries
which possess trapped surfaces and those which do not,
${\cal D}_0$. One would expect that ${\cal D}_0$ is bounded with respect
to any reasonable norm. A geometry can only be
singular if $R^\prime=0$ somewhere so that
it also possesses a trapped surface. That there is no
natural supermetric which discriminates against geometries possessing
an apparent horizon indicates the very different natures of the boundaries,
${\cal S}$ and ${\cal T}$.

\vfill\eject

\centerline{\bf 4. THE MOMENTARILY STATIC CONSTANT DENSITY STAR}
\vskip1pc
\noindent{\bf 4.1 The Exact Solution}
\vskip1pc
In general, we cannot solve Eq.(2.1) exactly for arbitrary
$\rho(\ell)$. What we can do is solve a few simple exactly-solved
problems consisting of piece-wise constant values of $\rho$.
Such models can clearly approximate a generic energy profile
arbitrarily closely. We can therefore exploit them as a guide to
identifying the generic features of solutions.

The simplest model there is is the constant density star.\footnote *
{We should point out that while the concept of a constant density does
not depend on the spatial system of coordinates it does depend on the
foliation. }
If $\rho$ is a constant in some interval $\ell \le \ell_0$ containing the
origin, we can recast Eqs.(2.5) and (2.6) as the single equation,

$$R^{\prime2}+{8\pi \rho\over 3}R^2  =1 \eqno(4.1)$$
in this region. Outside the support of matter,

$$R^{\prime2}+{2 m_\infty\over R} =1\,,\eqno(4.2)$$
where $m_\infty = {4\pi\over 3}\rho R(\ell_0)^3$
is the ADM mass. We note that both $R$ and $R^\prime$
are continuous across $\ell=\ell_0$.

We can analyse Eqs.(4.1) and (4.2) by drawing on the analogy with
the one dimensional motion of a particle ($\ell$ is time)
with unit `energy' in a time dependent potential, $V(R,\ell)$, which
changes from a harmonic oscillator to a Coulomb potential at $\ell=\ell_0$:

$$ V(R,\ell)= \cases{ {8\pi \rho\over 3}R^2 & $0\le \ell\le \ell_0$ \cr
{2 m_\infty\over R}&$\ell_0 \le \ell\,.$\cr}\eqno(4.3)$$
This analogy will be particularly useful below when we examine
less trivial models. In the present case, the particle starts
out at the origin $R=0$ at $\ell=0$ with unit `velocity'
$R^\prime=1$. The classical turning point marking the position of
an apparent horizon (if one exists) is given by $R^\prime=0$ or
$R_c^2=3/ (8\pi \rho)$.
This represents the Schwarzschild radius of the inner apparent horizon.
Inside matter, $R=R(\ell)$ is given by

$$\ell=\int_0^{R} {dR\over\sqrt{1-\left({R\over R_c}\right)}}
= R_c \sin^{-1} \left({R\over R_c}\right)\,,\eqno(4.4)$$
or $R=R_c\sin \left({\pi \ell\over 2 \ell_c}\right)\,.$
where $\ell_c$ is given by $\ell_c={\pi\over 2} R_c$.
This is simply the metric on a round three-sphere $S^3$ of radius $R_c$.
We note that the interior of the constant density star is an Einstein
geometry, ${\cal R}_{ab}={1\over 3} {\cal R} g_{ab}$.
Thus the non-singularity of ${\cal R}$ guarantees that of
${\cal R}_{ab}$. This is the unique such geometry. This is because
all Einstein manifolds must possess constant ${\cal R}$ in dimensions
greater than two (for otherwise it would be inconsistent with
the contracted Bianchi identities) and in $d=3$, the unique
spherically symmetric solution with positive ${\cal R}$
and with the given topology is part of the round $S^3$.

The qualitative behavior of the solution depends on the size of the star,
$\ell_0$.

\noindent If $\ell_0< \ell_c$, $R$ increases monotonically
both inside and out and there is no apparent horizon.

\vskip1pc

\noindent if $\ell_0=\ell_c$, an apparent horizon
forms on the surface of the star. $R$ is monotonic everywhere.
We note that $R^{\prime\prime}$ is discontinuous here.
Generically, we would expect both $R^\prime$ and
$R^{\prime\prime}$ to vanish at this bifurcation.

\vskip1pc

\noindent If $\ell_0 > \ell_c$, $R$ increases monotonically inside
the star until it reaches a maximum at the (inner)
apparent horizon at $\ell=\ell_c$ at which $R=R_c$.
Beyond $\ell_c$, $R$ begins to decrease inside the star.

\vskip1pc

\noindent If $\ell_0\ge 2\ell_c$, $R$ returns to zero at $\ell=2\ell_c$
inside the star at which point $R^\prime=-1$.
There is a bag of gold singularity contained within the support of matter.
This is the only way that the geometry can exhibit a singularity in the
uniform density model. The model does not therefore display
generic singular behavior. This is its principle shortcoming.

\vskip1pc

\noindent If $2\ell_c\ge \ell_0\ge \ell_c$, we exit the star before
a singularity is reached. However, $R^\prime$
is continuous at the boundary so that $R$ continues decreasing
outside the star until an exterior horizon is reached at a value of $R$
determined by $R^\prime=0$ in Eq.(4b) or $R= 2 m_\infty$.

\vskip1pc
We note that the constraints only admit single bubble non-singular
configurations. Any higher number of bubbles is necessarily singular.



\vskip1pc
\noindent{\bf 4.1.1 Trapped Surfaces and Singularities:
Necessary and Sufficient Conditions}
\vskip1pc

What is most interesting about this model is that it indicates that
there exist bounds on the product of the energy density
by the square of the maximum support of a given uniform distribution of
matter beyond which the associated geometry (i)
possesses a trapped surface and (ii) is singular.

The surface of the star first becomes trapped (an apparent
horizon forms) when $\ell_0=\pi R_c/2$ so that
$\rho \ell_0^2 = {3\pi\over 32}$. If
$\rho \ell_0^2 \ge {3\pi\over 32}$ then the star will
contain a trapped surface (either the surface of the
star will be trapped  or a trapped surface is contained
in its interior) and conversely.\footnote *
{We need not worry about trapped surfaces
in the region exterior to the star as they
will always be attended by an interior one.}

The surface of the star will be	singular when $\ell_0=\pi R_c/2$ or
$\rho \ell_0^2 = {3\pi\over 8}$. If $\rho \ell_0^2 \ge {3\pi\over 8}$
then the star will contain a (weak) singularity (either on its surface
or in its interior) and conversely. We summarize,

$$\rho \ell_0^2 \ge {3\pi\over 32}\quad {\rm iff\quad (i)}\quad;\quad
\rho \ell_0^2 \ge {3\pi\over 8}\quad {\rm iff\quad (ii)}\,.\eqno(4.5)$$
The necessary and sufficient conditions formulated at $\ell_0$
not only coincide but they also make a non-trivial statement
about the interior physics.

\vskip1pc
\noindent{\bf 4.2  $M$ vs. $\ell_0$,
$m$ vs. $\ell_0$ and the characterization of
trapped Surfaces and singularities in terms of these variables}
\vskip1pc

Unfortunately, the unambiguous partitions of the
configuration space represented by Eq.(4.5)
are a very special feature of the constant density star. If
the density is not constant, it will not be possible in general
to phrase both the necessary and the sufficient condition
with respect to the same set of variables. When it is possible,
the inequalities analogous to Eq.(4.5) typically will
differ. The if statement will be stronger than the only if statement.
In addition, our ability to make non-trivial statements about the
interior based on surface measurements will, in general, depend on the
choice of variables.

We need to identify appropriate variables to characterize the
material energy contained within a sphere of
fixed proper radius when the energy density is not constant.
Typically, this will be some (weighted) mean value of $\rho$
over the interior volume. The two measures we introduced in paper I
were the QLM $m$ and the material energy $M$.

The simplest global measure of the material energy content of a
spherically symmetric system is

$$M(\ell_0) = 4\pi \int_0^{\ell_0} d\ell R^2 \rho\,.\eqno(4.6)$$
Let us examine $M$ as a function of $\ell_0$ at constant $\rho$
in the constant density model. We have

$$M=4\pi\rho\int_0^{\ell_0} d\ell\, R^2=
{3\ell_0\over 4}\left[1- {R_c\over 2\ell_0}\sin
\left({2\ell_0\over R_c}\right)\right]\,.\eqno(4.7)$$
What is more relevant, however, is the ratio $M(\ell_0)/\ell_0$.
This is illustrated in Fig.(4.1).

In this model it is already evident that crucial information is lost
in replacing $\rho$ by $M$ to characterize the energy content of the star.
The function $\ell\to M(\ell_0)/\ell_0$ is neither surjective nor
injective --- it is bounded and (because the ratio oscillates) it is
not unique. The reason for this, as we will discuss further below,
is the folding of $R$ into the definition of $M$.

Let us attempt to formulate necessary and sufficient
conditions for the presence of trapped surfaces
and singularities in a constant density star using $M(\ell_0)$ as our
measure of material energy to see how it fares.
We should not be surprised that we are unable to do as well as
we did with $\rho$ --- $\rho$ and $\ell_0$ after all completely
characterize the star, whereas $M(\ell_0)$ and $\ell_0$ do not.

\vskip1pc
\noindent{\bf 4.2.1 Trapped Surfaces}
\vskip1pc

An apparent horizon forms at the surface of the star when
$M(\ell_0) = 3\ell_0 /4$.

\item{} Sufficiency: If

$$M(\ell_0)\ge 3\ell_0 /4\eqno(4.8)$$
the star will contain a trapped surface. This condition is sharp.
The surface itself need not be trapped. However, if it is
not then the interior must be singular and
consequently must contain an apparent horizon. So far so good.

The tightest inequality describing necessity is considerably weaker: If

$$
M(\ell_0)<    {3 \ell_0\over 4}(1 - {2\over 5\pi})\,,\eqno(4.9)$$
the star does not contain a trapped surface.
An important point is that the inequality (4.9) describing necessity
is weaker than the one describing sufficiency.
In fact, inspection of Fig.(4.1) indicates that this inequality
is not sharp and, as such, of limited value for there are stars with
$M(\ell_0)$ exceeding the value on the RHS of Eq.(4.9) without
any trapped surfaces. If it had made sense to demand that
the interior be non-singular in the hypothesis,
then, when $M(\ell_0) <3\ell_0/4$, the star would not contain a
trapped surface. The possibility of singularities
lurking in the interior beyond which the ratio $M(\ell_0)/\ell_0$
is decreasing is what really limits our ability to
formulate the necessary condition in terms of $M(\ell_0)$ and $\ell_0$.

For trapped surfaces, we can conclude that $M$ is a good
variable for the formulation of a sufficiency condition but not for the
formulation of a necesssary one.

\vskip1pc
\noindent{\bf 4.2.2 Singularities}
\vskip1pc

The surface of the star will be singular when the ratio of $M(\ell_0)$
to $\ell_0$ is $3/4$, which is the same as the apparent horizon ratio.
This first occurs, however, when $\ell_0=\pi R_c/2$ or
$\rho \ell_0^2 = {3\pi\over 8}$ which is different from
the apparent horizon value. The description in terms of
$\rho$ and $\ell_0$ clearly provides a more precise characterization.
When we attempt to provide necessary and sufficient conditions  on
$M(\ell_0)$ and $\ell_0$ to describe singularities we discover
unforeseen complications. Not only are we unable to formulate
conditions which coincide, we find that we cannot even formulate a
consistent (non-vacuous) sufficiency condition	of the form: if
$M(\ell_0)\ge \alpha \ell_0$ for some $\alpha$, then
the star will contain a singularity.
What we possess instead is a universal bound on the ratio

$$M(\ell_0)/\ell_0 \le  {3\over 4}(1 + {2\over 3\pi})\eqno(4.10)$$
which holds for all geometries which satisfy the constraints,
whether they are (weakly) singular or not.
The geometry which saturates the inequality is unique. It is the
configuration packing the maximum material energy into a given
proper radius. It is non-singular but does contain a trapped surface.

The only necessary condition we can provide with these variables
is the somewhat trivial one implied by the corresponding trapped
surface condition. If the star contains a singularity, it will also
contain a trapped surface. However, if this criterion was poor
for trapped surfaces, it is even poorer as a characterization
of singularities.

The reason why we are unable to provide a
sufficiency condition in terms of $M(\ell_0)$ and $\ell_0$
which is satisfied by an non-empty set
is subtle and not only a reflection of the inadequacy of these
variables to describe non-singular geometries. We have already noted
that a star with a monotonically decreasing $\rho$ cannot be
strongly singular. If the decrease in $\rho$ is strict, neither can it
be even weakly singular. We will demonstrate in Sect.6 that
a sufficient condition for the presence of a strong singularity
is $M(\ell_0)>2\ell_0$. We can conclude that for all geometries with
$\rho^\prime \le 0$, $M(\ell_0)\le 2\ell_0$.\footnote *
{In Sect. 7, we will show that we can do better.}
The sufficiency condition provides a universal bound on this subset
of the configuration space. The existence of the bound is not simply
an artifact of the constant density model.

In this regard, we should stress another distinction between the
choice of $\rho$ and that of $M(\ell_0)$ as a measure of the
material energy which is that with $\rho$ one
can distinguish between weakly singular and
non-singular geometries whereas with $M(\ell_0)$ one cannot.
The sufficiency condition, $\rho\ell_0^2\ge 3\pi/8$ is,
however, unstable. For if we subject the constant density
to a strictly monotonically decreasing perturbation,
the subset satisfying this condition also becomes empty.

\vskip1pc
\noindent{\bf 4.2.3 $m(\ell_0)$ vs. $\ell_0$}
\vskip1pc

If $M(\ell_0)$ fares poorly for the purpose of characterizing
the interior geometry, the QLM $m(\ell_0)$ will fare even more poorly
(see Sect.7 of paper I). To demonstrate this explicitly,
let us compare the analytical form for $M$, Eq.(4.7), with that for $m$. We
have

$$m(\ell_0)={4\pi\over3} \rho R(\ell_0)^3
={R_c\over 2}\sin^3 {\ell_0\over R_c}\,.\eqno(4.11)$$
$m(\ell_0)$ is bounded for a given $\rho$,
$m\le {R_c\over 2} =\sqrt{3/(32\pi \rho)}$,
assuming its maximum when a horizon forms at $\ell_0$, decreasing
thereafter as a function of $\ell_0$ as matter is deposited
behind the outer horizon. It falls to zero with the appearance of a (weak)
singularity. There are two non-singular configurations corresponding to each
value of $m(\ell_0)$ --- one with, the other without an apparent
horizon.

\vskip1pc
\noindent{\bf 4.3 Binding Energy in the constant density star}
\vskip1pc

In Sect.7 of paper I, we saw that the `binding energy',
$m-M$ is always negative. In the weak field limit, we note that

$$M\sim {1\over 2}{\ell_0^3\over R_c^2} -
{1\over 10}{\ell_0^5\over R_c^4}+\cdots\,,\eqno(4.12a)$$
while

$$m\sim {1\over2}{\ell_0^3\over R_c^2} -{1\over 4}
{\ell_0^5\over R_c^4}+\cdots\,.\eqno(4.12b)$$
The leading cubic terms in $\ell_0$ in both coincide. It is just
the `rest' mass, $m_0=4\pi\rho \ell_0^3/3$ associated with a uniform
spherically symmetric distribution of matter in special relativity.
Thus to leading order, the difference is given by

$$m-M\sim
- {3\over 5} {M^2\over \ell_0}\eqno(4.13)$$
which is the Newtonian binding energy of a spherical uniform density star.
We note that at the other extreme where the geometry is about to
turn singular, $\ell_0\sim \pi R_c$, $m=0$ and
$M\sim {4\over 3} M^2/\ell_0$.
We note that in any non-singular geometry the binding energy is
bounded from below as follows [10] (see Sect.7 below),

$$M-m\ge {1\over 2} {M^2\over \ell_0}\,.$$
For a constant density star, there is the tighter lower bound,
$M-m\ge {3\over 5} M^2/\ell_0$.
The coefficient on the RHS increases monotonically with $\ell_0$. We
conjecture that whenever $\rho^\prime \le 0$ this inequality will hold.

\vfill
\eject

\centerline{\bf 5. MSCs OF THE 2-PIECEWISE CONSTANT DENSITY STAR }
\vskip1pc

The constant density model lies on the boundary of the
subset of montonically decreasing $\rho$, admitting weakly but not
strongly singular geometries. Such weak singularities occur with
zero measure in the set of all possible singular geometries.
The constancy of $\rho$ provides very strong global information on the
surface concerning the interior.

While it is a very useful model to get us started, it is
important to recognize these limitations in order
to avoid making erroneous extrapolations. These limitations
will be highlighted further by examining a slighly more complicated model.

A concrete model which does display generic behavior
is a star consisting of an inner spherical ball of
radius $\ell_1$ with energy density $\rho_1$ and an outer shell of
thickness $\ell_2-\ell_1$ with energy density $\rho_2$:

$$\rho(\ell)=\cases{\rho_1& $0\le \ell \le \ell_1$\cr
\rho_2& $\ell_1 < \ell \le \ell_2$\cr
0&$\ell_2 < \ell\,.$\cr}\eqno(5.1)$$
In the region $\ell\le \ell_1$, the solution is just one of the
solutions we obtained in our examination of the constant density star.
In $\ell_1 < \ell \le \ell_2$,

$$R^{\prime2} +V(R) =1\,,\eqno(5.2)$$
where

$$V(R) = {2\Delta m\over R} +{8\pi \rho_2\over 3}R^2\,,\eqno(5.3)$$

$$\Delta m= {4\pi\over 3}R^3_1 (\rho_1 -\rho_2)\,,\eqno(5.4)$$

$$R_1= R_{c1}\sin \left({\ell_1\over R_{c1}}\right)\,,\eqno(5.5)$$
and $R_{c1}^2= 3/( 8\pi \rho_1)$.
$\Delta m$ is a measure of the energy density excess in the interior.

As we will see, the qualitative behavior of the solution in the
region $\ell_1 < \ell \le \ell_2$ depends very crucially  on the
sign of $\Delta m$. Let us therefore examine the two signs one at a
time ($\Delta m=0$ is trivial). We will suppose that the interior
geometry is regular in both cases.

\vskip1pc
\noindent{\bf 5.1 $\Delta m > 0$}
\vskip1pc

Let us first suppose that $\Delta m > 0$.
When $\Delta m >0$, the potential $V(R)$ is bounded
from below with a positive minimum at $R_m$ given by

$$R_m= \left({3\Delta m\over 8\pi\rho_2}\right)^{1/3}\,.\eqno(5.1.1)$$
$V(R)=1$ possesses two positive roots and $V(R_m)< 1$ whenever $\Delta m>0$.
This confirms that there is no critical positive value of $\Delta m$
beyond which no solution exists. One way to show this is to note that
$V(R_m)= 3{\Delta m \over R_m}$ so that $V(R_m)\ge 1$ implies
$\Delta m^2 \ge {1\over 72\pi \rho_2}$. If we define
$x={\rho_2\over \rho_1}$, the
inequality $V(R_c)\ge 1$ can be expressed in the dimensionless form

$$\sin^3
\left({\ell_1\over R_{c1}}\right)\,(1- x)^2 \ge {4\over 27 }{1\over x}\,.
\eqno(5.1.2)$$
Now, if $\Delta m>0$ then  $x <1$  which is inconsistent with Eq.(5.1.2).

The most important feature displayed by this solution is that the
geometry cannot be singular on the interval $\ell_1 < \ell \le \ell_2$
or outside the star. The `potential' provides a barrier keeping the solution
away from $R=0$. No matter how large the value of $\ell_2$, the solution is
periodic and non-singular on the interval $\ell_1 < \ell \le \ell_2$
oscillating between turning points given by the two positive roots of
$V(R)=1$. This provides a useful model of a
(strictly somewhere) monotonically decreasing energy density.
This should be contrasted with the uniform density star
which turns singular beyond some critical value of
$\ell_0$. The inner higher density region has a stabilizing effect
on the geometry of the outer region.

We note that the region $\ell\le\ell_1$ can be viewed as a momentarily
static slice through de Sitter space. The geometry in the
interval $\ell_1 < \ell \le\ell_2$ is the momentarily static slice
through a Schwarzschild-de Sitter spacetime. If $\rho_1$ and $\rho_2$
are both constant in time, this initial data will generate these
spacetimes in their respective intervals. There will be, in fact,
an infinite number of such geometries with the same ADM mass which we
can label by the number of maxima of $R$. The existence of such geometries
might have implications in semi-classical quantum gravity.
The semi-classical calculation of tunneling amplitudes
involves a summation over $\exp -S$, where $S$ is the classical action
evaluated on the Euclidean interpolation from one
MSC to one of the infinite number of MSCs with the same ADM mass.
This sum could potentially result in a non-trivial
amplification of the transition amplitude.

There is a theorem due to Schoen and Yau stating that if the
volume (defined in terms of a certain embedded torus) supporting a given
strictly positive $\rho$ is made large in all directions, the geometry
must be singular [14]. It would appear that our ability to make
$\ell_2$ arbitrarily large without incurring any singularity is inconsistent
with this result. However, a closer examination shows that this is not so.
This is because even though the proper volume associated with
a large value of $\ell$ is large (increasing linearly with $\ell$),
it is not large in all directions. It is essentially a cylinder of
bounded radius. Thus, a large torus in the sense of Schoen and Yau
does not fit.

\vskip1pc
\noindent{\bf 5.2 $\Delta m <0$}
\vskip1pc

What is more dramatic is the geometry which corresponds to $\Delta m<0$.
The potential is now monotonic and unbounded from below within the
interval $\ell_1 < \ell\le \ell_2$.

\noindent If $\ell_2 < \ell_s$, where $\ell_s$ is the proper radius at
which $V(R(\ell_s))=0$ the geometry will be regular both inside and out.
Let $R_s=R(\ell_s)$. $R_s$ is given by

$$R_s=\left({3\Delta m\over 4\pi\rho_2}\right)^{1/3}\,,\eqno(5.2.1)$$
and

$$\ell_s=\ell_1 +\int_{R_1}^{R_s} {dR\over \sqrt{1- V(R)}}\,,\eqno(5.2.2)$$
where $V(R)$ is given by (5.3), (5.4) and $R_1$ is given by Eq.(5.5).
The ADM mass, $m_\infty= m(\ell_2)$. Therefore

$${2 m_\infty\over R_2} =
{2\Delta m\over R_2} +{8\pi \rho_2\over 3}R_2^2\,.\eqno(5.2.3)$$
When $\ell=\ell_s$, $R^\prime=-1$ and the QLM vanishes, $m(\ell_s)=0$.
If $\ell_2=\ell_s$ the geometry is flat outside. However,
before it can become asymptotically flat,
the geometry must close in a bag of gold. The empty top of the
bag consists of a flat cap (in which $R^\prime=-1$).
$R(\ell)=0$ at

$$\ell_{S}=\ell_s+ R_s\,.\eqno(5.2.4)$$
This is the same (non-generic) kind of singularity we
encountered in the uniform density model.

Recall that a regular closed bag requires the integrability
condition $m(\ell_S)=0$ to be satisfied. A necessary condition is
that $\rho$ assume a maximum (or a minimum) somewhere.
The simplest realization of such a configuration involves three
constant density slabs with $\rho_1<\rho_2>\rho_3$ (or
$\rho_1>\rho_2<\rho_3$). To satisfy the integrability
condition we will be required to tune the relative values
of these densities and their supports appropriately.
The model we are considering with $\rho_1<\rho_2$ with a vacuum
exterior represents a special case. The value $\ell_{S}$ given by Eq.(5.2.4)
is the unique critical value of $\ell_0$ supporting an everywhere
regular closed cosmology for the given parameter values.

\noindent If $\ell_2> \ell_s$, $R^\prime <-1$ at $\ell=\ell_2$
and $m(\ell)$ will be negative for all subsequent $\ell$ until
a singularity is reached at some point $\ell_S$.
If, in addition,  $\ell_2 \ge \ell_S$ where

$$\ell_S=\ell_s +\int_0^{R_s} {dR\over \sqrt{1- V(R)}}$$
is finite, the geometry closes with a metric singularity
$R^\prime\to -\infty$ before we ever reach $\ell_2$.
In the neighborhood of $R=0$, $R$ satisfies Eq.(2.8)
with singular solution (2.9). The local value of the QLM
at the singularity is given by $m(\ell_S)=\Delta m$.
Note that the quasilocal mass at the singularity is finite
and determined completely by the remote behavior
of the sources. If $\ell_2 <\ell_S$, the
interior is fitted with a negative $m$ exterior vacuum
cap with the same singular behavior.

\vfill\eject

\noindent{\bf 6. APPROXIMATING THE BOUNDARIES IN ${\cal C}$}

\vskip1pc
\noindent{\bf 6.1 Inequalities of Sufficiency and Necessity}
\vskip1pc

It is clear that a generic energy profile can, in principle,
be approximated arbitrarily closely by a a piece-wise
constant density model along the lines of our examination
of the two simplest such models in Sects. 4 and 5.
Even still, however, it is not obvious how
to exploit such models to characterize the boundaries partitioning
${\cal C}$ which were discussed in Sect.3. The approach we will take
is to establish
necessary and sufficient conditions determining the circumstances
under which the geometry can possess trapped surfaces or singularities [8,
9]. If we could solve the problem exactly,
as we did on the piece-wise constant density subset, presumably
the necessity and sufficiency conditions could
be formulated as if and only if statements with an appropriate
choice of variables.\footnote * {Of course, one could
argue that whenever the problem is exactly solvable,
in principle at least, one could simply tick off geometries
with $R'=0$ or $R=0$ at some finite value of $\ell$. We do not consider
this as a solution to the problem because it is telling us
nothing about the physics conspiring to produce these
coincidences.}

Let us examine in very general terms how closely these necessary
and sufficient conditions identify the (singular / non-singular say)
partition ${\cal S}$ of ${\cal C}$. The necessary condition will
provide us with a surface ${\cal S}_{\rm nec}$
bounding some region in the configuration space ${\cal C}_{\rm nec}$
interior to which we are assured that the geometry is non-singular.
The sufficiency condition will provide us with some other
surface ${\cal S}_{\rm suff}$ bounding a region in the configuration
space ${\cal C}_{\rm suff}$ exterior to which the geometry is always
singular. ${\cal C}_{\rm nec}$ will be some proper subset
of ${\cal C}_{\rm suff}$. The set difference ${\cal C}_{\rm suff}-
{\cal C}_{\rm nec}$ will not be empty, representing a grey area
containing both singular and non-singular geometries.
As the constant density model illustrates,
this discrepancy might be a consequence not only
of our inability to solve the problem exactly but also of our
inability to identify an optimal set of variables.

Ideally, we would like both the necessary and the sufficient conditions to
be sharp, namely that a configuration exists saturating the condition,
though we do not necessarily expect the two to be sharp simultaneously.

\vskip1pc
\noindent{\bf 6.2 Sufficiency and Necessity in terms of $M$ and $\ell_0$}
\vskip1pc

Let us rewrite Eq.(2.1) in the form

$$4\pi  \rho R^2  + \big(RR^\prime\Big)^\prime=
{1\over 2}\Big(1+(R^\prime)^2\Big)\,,\eqno(6.2.1)$$
and integrate out from $\ell=0$ up to the surface value $\ell_0$:

$$M (\ell_0) + RR^\prime\Big|_{\ell_0}= \Gamma\,,\eqno(6.2.2)$$
where we define

$$\Gamma \equiv {1\over2}\int_0^{\ell_0} d\ell (1+(R^\prime)^2)\,.
\eqno(6.2.3)$$
We now have two different representations for the
first integral of the constraints. Earlier we exploited the expression
Eq.(2.5)/(2.6) which involves the definition of the QLM.
Eq.(6.2.2) involves $M(\ell_0)$ and will provide the basis for our
derivations of all inequalities involving
$M$.

\vskip1pc
\noindent{\bf 6.2a Trapped Surfaces: Sufficiency}
\vskip1pc

In our examination of the constant density star, we found that
the sufficient condition for the presence of a of trapped surface
is as good as one could hope for --- the inequality (4.11) is sharp.
This success is repeated for generic density profiles.

We begin by eliminating the term involving  $R^\prime(\ell_0)$ from
Eq.(6.2.2) in favor of the expansion of the optical scalar
$\Theta_-=\Theta=\Theta_+$ (see appendix of I):
$\Theta =2R'/R$. If the surface is not trapped, $\Theta(\ell_0)>0$ and thus
$M(\ell_0) < \Gamma$. We now require an upper bound on $\Gamma$. When
$\rho\ge0$ and the interior is regular, then
$(R^\prime)^2\le 1$ so that

$$\Gamma \le \ell_0\,.\eqno(6.2.4)$$
Thus

$$M(\ell_0) < \ell_0\,.\eqno(6.2.5)$$
In other words, if
$M(\ell_0) \ge \ell_0$ the surface must be trapped or the
interior must be singular in which case it also
possesses a trapped surface.\footnote * {Note that in the
derivation, we did not have to assume that
the interior did not contain a trapped surface.}
BMO'M have demonstrated that this inequality is sharp. They do this by
constructing a model in which $M(\ell_0)=\ell_0(1-\epsilon)$ but the
surface is not trapped.

It is possible to tighten the sufficiency
condition whenever $\rho^\prime\le 0$.
We exploit Eq.(2.5) to express $\Gamma$ as a functional of $m(\ell)$:

$$\eqalign{\Gamma =&{1\over2}\int_0^{\ell_0} d\ell
(1+(R^\prime)^2)\cr
=&\int_0^{\ell_0} d\ell \left[ 1-{1\over 2}(1-(R^\prime)^2)\right]\cr
=&\int_0^{\ell_0} d\ell \left[1-{m(\ell)\over R}\right]\,.\cr}\eqno(6.2.6)$$
We now exploit Eq.(2.10) which holds whenever
$\rho^\prime\le 0$ to obtain

$$\Gamma \le \ell_0-{4\pi\over3} \int_0^{\ell_0} d\ell \rho R^2
= \ell_0 -{M(\ell_0)\over 3}\,,\eqno(6.2.7)$$
which is a tighter bound on $\Gamma$.
Thus, in place of Eq.(6.2.5) we have

$${4\over 3} M(\ell_0) < \ell_0\,.\eqno(6.2.5^\prime)$$
Thus, if $M(\ell_0) \ge  {3\over 4}\ell_0 $ the surface of the monotonic
configuration must be trapped or it contains a trapped surface.

A nice feature of Eq.(6.2.5$^\prime$) is that not only is it sharp,
but we also know that it is saturated by the constant density star.
In addition we note that it is easier to form a trapped surface
when $\rho$ is monotonically decreasing.
This is to be contrasted with the impossibility of finding a
singularity when $\rho$ is monotonically decreasing.
If we scale the value of $\rho$ or its support, a
critical point will always be reached at which a trapped surface forms.
However, there is no corresponding point marking the formation of a
singular geometry. This exposes the inadequacy of the
intuitively reasonable notion that by increasing $\rho$ on a constant support
we first produce a trapped surface and then a singularity.

We note that the sufficiency condition can be cast in a
weaker form in terms of the minimum energy density
$\rho_{\rm Min}$ using
$\rho_{\rm Min} V_0\le M(\ell_0)$ where $V_0$ is the volume of the star.
In the monotonic star, this will be slightly more useful on
account of the fact that $\rho_{\rm Min}$ will be the surface energy
density. More useful inequalities involving $\rho_{\rm Min}$ can be
derived by exploiting the calculus of variations in the manner of
Schoen and Yau [14].

\vskip1pc
\noindent{\bf 6.2b Trapped Surfaces: Necessity}
\vskip1pc

In the constant density star, we found that $M$ was
a poor variable for formulating necessary conditions for the
presence of trapped surfaces and of singularities.

In general, if the surface
is trapped, $\Theta(\ell_0)\le 0$. Then Eq.(6.2.2) implies that
$M(\ell_0) \ge \Gamma$. We now require a lower bound on $\Gamma$.
The weak lower bound on $\Gamma$,
$\Gamma \ge {\ell_0\over 2}$,
relies only on the fact that $(R^\prime)^2\ge 0$ which
is independent of the positivity of $\rho$. We get
$M(\ell_0)\ge {\ell_0\over 2}$ on a trapped surface. Thus if

$$M(\ell_0) <{\ell_0\over 2} \eqno(6.2.6)$$
the surface is not trapped. It is not sharp and it fails
to prevent a trapped surface showing up
in the interior. Compare Eq.(6.2.6) with the marginally better
Eq.(4.9) we obtained for the
constant density model which, in addition,
guaranteed that no trapped surfaces showed up
in the interior. The reason for the latter stronger statement
is the fact that we were able to exploit the exact solution to
feed global information into the surface condition.
As an extreme example, consider a star with a
high density core of radius $\ell_1$ and a large low density
mantle. We might then have $M(\ell_1)\ge\ell_1$
but $M(\ell_0) \le\ell_0/2$ on the surface. In other words, the
surface information  provides no clue as to the interior physics.
For this reason the condition $\rho^\prime < 0$ does not allow us
to `tighten' the necessary condition the way it did the
sufficiency condition. Technically, the inequality is
operating the wrong way.
What is possible is to tighten the
necessary condition in a configuration with a radially
increasing energy density $\rho^\prime\ge 0$ to

$$M(\ell_0) \le 3\ell_0/5\,.\eqno(6.2.6')$$
This is the opposite extreme to the central high density core.
It is harder to produce a trapped surface.
However, even on this subset, we still cannot provide a strong
form of necessary condition prohibiting trapped surfaces
showing up in the interior. We note that the constant density star is
also a special case of Eq.(6.2.6$'$).  Eq.(6.2.6$'$)
is not sharp because Eq.(4.9) is not.
We will return to the issue of formulating
necessary conditions below using $\rho_{\rm Max}$ instead of $M$.

We will find a necessary condition analogous
to the constant density result holds when we use
$\rho_{\rm Max}$ instead of $M$. The reason is that $\rho_{\rm Max}$
taps into global information without the distortions introduced by folding
$R$ into the definition of $M$.

\vskip1pc
\noindent{\bf 6.2c Singular Geometries: Inequality of Sufficiency}
\vskip1pc

Because $M$ does not discriminate between the regular and the weakly
singular geometries exhibited by the constant density model, we
found that instead of a sufficiency condition for
the presence of singularities, the inequality (4.10)
represented a universal bound.

In general $M$ does discriminate between geometries which
are strongly singular and those which are not. Therefore it can provide
a non vacuous sufficiency condition for the presence of singularities.
This bound was obtained by BM\'OM [8,9].
On the monotonically decreasing density subset, it again provides
a universal bound though not a sharp one.

To reproduce the sufficiency condition, let us suppose that the geometry
is not strongly singular. Then $R^{\prime2}\le 1$, so that
$\Gamma\le \ell_0$ and Eq.(6.2.2) implies

$$M (\ell_0) + RR^\prime\Big|_{\ell_0} \le \ell_0\,.\eqno(6.2.7)$$
In addition, $R(\ell)\le \ell$ everywhere and $R^\prime\ge -1$. The
surface term is therefore bounded from below by $-\ell_0$ and

$$ M(\ell_0) \le 2\ell_0\,.\eqno(6.2.8)$$
If, in particular $\rho^\prime\le 0$, then this condition
represents a  universal bound on $M$ in a MSC.
It is not possible in
such a model to raise the value of $M$ indefinitely by increasing
$\rho$ while maintaining $\ell_0$ constant.
What happens when we do this is that $\rho R^2_{\rm Max} $ saturates
within the star and as
a consequence the increase in $M(\ell_0)$ is at most linear in $\ell_0$.
We will see how this occurs explicitly below in the context of a
two density model. In a large outer low density region
$R$ does not grow with $\ell_0$ but oscillates
between fixed mimimum and  maximum values. This is folded into the
definition of $M$.

We can tighten the inequality whenever $\rho^\prime\le 0$
just as we did the trapped surface sufficiency condition.
Let us do it in a slightly different way which exploits
the  Schwarzschild inequality explicitly.
The Schwarzschild radius, $R(\ell)$,
provides a bound on the quasilocal mass
$2 m(\ell)\le R(\ell)$
whenever the constraints are satisfied  (whether the geometry is
singular or not).
This is a straightforward consequence of Eq.(2.5).

$$m-{R\over 2}=-\left(R^\prime\right)^2 {R\over2 }\,.$$
The right hand side is bounded above by zero ($R^{\prime2}\ge 0$).
Thus

$$m-{R\over 2}\le 0\,.\eqno(6.2.9)$$
Equality obtains on an apparent horizon $R^\prime=0$
and when $R=0$ where $m=0$.

The condition (2.8) and the Schwarzschild inequality
(6.2.9) together then imply that
$\rho R^2 \le 3/ (8\pi)$ inside matter. Thus

$$M(\ell_0)=4\pi\int_0^{\ell_0} d\ell \,R^2 \rho
\le  {3\over 2} \ell_0\,.$$
It is curious that, unlike the corresponding trapped surface condition,
this inequality is not sharp.

\vskip1pc
\noindent{\bf 6.2d Singular Geometries: Necessity}
\vskip1pc
Suppose that the geometry possesses a singular surface at $\ell_0$.
Then $R(\ell_0)=0$.
What's more the product
$R R^\prime$ appearing in Eq.(6.2.2)
also vanishes at a singularity even though
$R^\prime$ itself might diverge. Thus

$$M(\ell_0) =\Gamma \ge {\ell_0\over 2}\eqno(6.2.9)$$
which is the same as the necessary condition for the
formation of trapped surfaces. This condition is
weak --- it does not even demand the positivity of $\rho$.
Neither does it say anything about the interior.

The necessary conditions we have obtained
are far weaker statements than we would like,
failing to eliminate the possibility of trapped surfaces
or singularities lurking in the interior.
What is clear by now is that $M(\ell_0)$ and
$\ell_0$ are not the ideal variables for this purpose.

We turn now to the task of formulating alternative
necessary conditions.

\vskip1pc
\noindent{\bf 6.3 Necessity in terms of
$\rho_{\rm Max}$ and $\ell_0$}
\vskip1pc

In the context of the constant density star we discovered that
the inequalities which describe the trapped/untrapped,
singular/non-singular partitions of geometries
is expressed with greater precision in terms of $\rho$ and $\ell$,
the more so for the inequalities of necessity.
One reason why this is so is that the single constant value
encodes global information.  However, another
reason is that the inequalities do not involve $R$.
What therefore are the appropriate generalizations of
the constant $\rho$ which are
independent of $R$?

The generalization which permits us to
formulate necessary conditions is the maximum value of $\rho$
on the support of matter, $\rho_{\rm Max}$.
The dimensionless combination of $\rho_{\rm Max}$ and
$\ell_0$ is $\rho_{\rm Max}\ell_0^2$. Such conditions
can be cast in the form: trapped/singular on surface or in the
interior implies
$\rho_{\rm Max}\ell_0^2 \ge $ some constant. What's more
the constants will be different in the two cases just as
they were for a constant density star.  Not only are these necessary
conditions stronger than those
obtained earlier in terms of $M$ for characterizing the
surface, they also detect trapped surfaces/singularities in the
interior if they are present.

\vskip1pc
\noindent{\bf 6.3a Singular Geometries}
\vskip1pc

If the interior geometry is singular then $R(\ell_1)=0$ at some $\ell_0\ge
\ell_1>0$.
We integrate Eq.(6.2.1) from $\ell=0$ up to $\ell_1$.
The integral over the divergence results in a surface term which
vanishes due to the boundary condition on $R$ at $\ell_1$. We obtain

$$\int_0^{\ell_1} d\ell (1+R^{\prime2})=8\pi\int_0^{\ell_1} d\ell
\rho R^2 \,.\eqno(6.3.1)$$
It is crucial for the successful implementation of the
integration that the product $R R^\prime$ vanishes at a singularity
even though $R^\prime$ itself might diverge. We now apply the
H\"older inequality with the $L^\infty$ norm on $\rho$ to obtain the bound

$$\int_0^{\ell_1} d\ell
\rho R^2 \le \rho_{\rm Max} \int_0^{\ell_1} d\ell
R^2 \eqno(6.3.2)$$
on the left-hand side. The bound is clearly exact if $\rho$ is constant
but is a weak estimate if $\rho$ varies a lot. Hence we get

$$\int_0^{\ell_1} d\ell (1+R^{\prime2})\le 8\pi\rho_{\rm Max}\int_0^{\ell_1}
d\ell  R^2
\,.\eqno(6.3.3)$$
The nice thing about this expression is that we can exploit
a  Sobolev inequality to get a second bound on the quantities occuring
in it.  For functions on the finite interval $(0,\ell_1)$
which vanish at the endpoints, there is a positive constant $S_0$
such that

$$S_0 \int_0^{\ell_1} d\ell
R^2  \le \int_0^{\ell_1} d\ell R^{\prime2}
\,.\eqno(6.3.4)$$
The inequality is saturated by  the trigonometric function
$R(\ell)=\sin (\pi \ell/\ell_1)$
which also determines the optimal value of $S_0$:
$S_0=\pi^2/\ell_1^2$.
$S_0$ is just the ground state energy of a quantum
mechanical particle in a box of width, $\ell_1$. In a constant density
star $R(\ell)$ is exactly of this form so that the second estimate
involved here is also exact. Substituting Eq.(6.3.4) in (6.3.3) we get

$$\int_0^{\ell_1} d\ell = \ell_1 \le (8\pi\rho_{\rm Max} -
{\pi^2\over \ell_1^2})\int_0^{\ell_1} d\ell R^2
\,.\eqno(6.3.5)$$
While we do not have a lower bound for $R^\prime$, we do have an upper
bound which is that $R^\prime \le 1$. This then gives us that $R \le
\ell$ and that

$$\int_0^{\ell_1} d\ell R^2 < {\ell_1^3 \over 3}\,.\eqno(6.3.6)$$
When this is substituted into Eq.(6.3.5) we get

$$\rho_{\rm Max} \ell_1^2 > {\pi \over 8} + {3 \over 8\pi}\,.\eqno(6.3.7)$$
Hence we conclude that if
$\rho_{\rm Max} \ell_1^2 < \pi/ 8 + 3/
(8\pi)$ we will not reach a singularity within a proper
distance $\ell_1$ from the origin.
This estimate is only one of a family of such estimates that can
be derived. Let us begin by multiplying Eq.(6.2.1) across by $R^{a}$
where $a$ is some, as yet undetermined, constant. We can write
(6.2.1) as

$$R^{a} +  (2a + 1)R^{a}(R^\prime)^2 - 2(R^{a + 1}
R^{\prime})^{\prime} = 8\pi\rho R^{a + 2}
\,.\eqno(6.3.8)$$
We integrate this out to a singularity at $\ell$ at $\ell_1$. The
divergence gives us a surface term which will vanish so long as $a >
-{1 \over 2}$. Thus we get

$$\int_0^{\ell_1} d\ell R^{a} + {4(2a + 1) \over (a + 2)^2}
\int_0^{\ell_1} d\ell ([R^{{a + 2\over 2}}]^\prime)^2 =
8 \pi\int_0^{\ell_1} d\ell \rho R^{a + 2}
\,. \eqno(6.3.9)$$
We now repeat the argument as above, replace $\rho$ with $\rho_{\rm Max}$
and use the Sobolev inequality (with the same constant) to get
$$\int_0^{\ell_1}
d\ell ([R^{{a + 2\over 2}}]^\prime)^2 \ge {\pi^2\over \ell_1^2}
\int_0^{\ell_1} d\ell \, R^{a + 2}
\,. \eqno(6.3.10)$$
When these are substituted into Eq.(6.3.9) we get

$$\int_0^{\ell_1} d\ell R^{a} \le \left[8\pi\rho_{\rm Max} -
{4(2a + 1)\pi^2 \over (a + 2)^2\ell_1^2}\right]
\int_0^{\ell_1} d\ell \, R^{a + 2}
\,. \eqno(6.3.11)$$

The last estimate needed is a bound on the ratio

$$\int_0^{\ell_1} d\ell\, R^{a + 2} \Big/
\int_0^{\ell_1} d\ell R^{a}\,.\eqno(6.3.12)$$
The very crude bound, $\ell_1^2 $, follows immediately from the fact that
$R^\prime \le 1$ which gives us that $R < \ell_1$. Because $R$ appears in the
denominator, we cannot naively exploit the bound $R\le \ell$ as we could in
Eq.(6.3.6). One can, however, show that the linear function $R = \ell$
maximizes the ratio (6.3.12) for all functions which satisfy the boundary
condition $R(0)=0$, and the two constraints $R^\prime\le 1$ and $R\ge 0$
(see appendix II). Thus

$${\int_0^{\ell_1} d\ell \, R^{a + 2} \over
\int_0^{\ell_1} d\ell\, R^{a}} < {1+ a \over3+ a}\ell_1^2
\,, \eqno(6.3.13)$$
holds. We thus obtain

$$\rho_{Max}\ell_1^2 \ge {(1 + 2a)\over (2 + a)^2}{\pi\over 2} +
{3 + a \over 1 + a}{1\over 8\pi}
\,.\eqno(6.3.14)$$
The dominant term on the RHS is the first.
The best value for it occurs when $a=1$. We get

$$\rho_{\rm Max} \ell_1^2 > {\pi \over 6} + {1 \over 4\pi}
\,.\eqno(6.3.15)$$
The RHS of Eq.(6.3.15) is slightly larger than
that of Eq.(6.3.7). Note that $a=1$
corresponds to an integrand $R R^{\prime2}$ which tends exactly
to a constant at a singularity.
More accurately, the RHS of (6.3.14) is maximizes when
$a \sim 0.8$. This is less than 1\% better than
Eq.(6.3.15). None of these numbers are
particularly impressive, they are, at best, only
half the value we get in the constant density star.

Despite our effort, it is clear that the estimate
Eq.(6.3.13) is the weakest link in the argument.
\vskip1pc
\noindent{\bf 6.3b Apparent horizons}
\vskip1pc

Suppose now that the geometry possesses an apparent
horizon, $R^\prime(\ell_1)=0$ at some $\ell_0\ge \ell_1>0$.
We again integrate Eq.(5.2) from $\ell=0$ up to $\ell_1$:
As before, the boundary condition kills the surface term and we are left with
an identical equation:

$$\int_0^{\ell_1} d\ell (1+R^{\prime2})
=8\pi \int_0^{\ell_1} d\ell \rho R^2
\,.\eqno(6.3.16)$$
The first approximation proceeds identically

$$\int_0^{\ell_1} d\ell (1+R^{\prime2})\le 8\pi
\rho_{\rm Max} \int_0^{\ell_1} d\ell R^2
\,.\eqno(6.3.17)$$
For the second approximation, we again exploit a Sobolev
inequality involving an appropriate function space on the right hand side.
If $R(\ell)$ is a function on the finite interval $(0,\ell_1)$
such that $R(0)=0$ and $R^\prime(\ell_1)=0$, then there is
some constant $S_1$ such that

$$S_1 \int_0^{\ell_1} d\ell R^2 d\ell
\le \int_0^{\ell_1} d\ell R^{\prime2}
\,.\eqno(6.3.18)$$
The inequality is again saturated by a sine function,
$R(\ell)=\sin (\pi \ell/2\ell_1)$ giving $S_1={\pi^2\over 4\ell_1^2}$.
In the case of a constant density star, both approximations are exact.
So

$$\int_0^{\ell_1} d\ell = \ell_1 \le (8\pi\rho_{\rm Max} -
{\pi^2\over 4\ell_1^2})\int_0^{\ell_1} d\ell R^2
\,.\eqno(6.3.19)$$
We again use Eq.(6.3.6),
and we finally get

$$\rho_{\rm Max} \ell_1^2 > {\pi \over 32} + {3 \over 8\pi}\,.\eqno(6.3.20)$$
This is a much better estimate than the corresponding
bound (6.3.7) or its improvements for the singularity, largely due
to the fact that (6.3.6) is much sharper in this case. In fact it is about
3/4 of the number that we get in the constant density star.

An expression equivalent to Eq.(6.3.14) can also be written down.
We have that

$$\rho_{Max}\ell_1^2 \ge
{(1 + 2a)\over (2 + a)^2} {\pi\over 8} +
{1\over 8\pi} {3 + a \over 1 + a}\,.\eqno(6.3.21)$$
The maximum of this expression occurs at about $a =
0.25$ and this improves on the bound Eq.(6.3.20) again by
about 1\%.

If $\rho$ is bounded from below within the star, its
minimum value, $\rho_{\rm Min}$ can be exploited to provide
alternative sufficiency conditions. However, if $\rho$ vanishes
anywhere this fails. In addition, because the inequalities of
sufficiency are derived by asssuming that the star does not possess a
trapped surface/singularity we cannot exploit Sobolev inequalities
in the way did for the necessary conditions.

\vfill\eject
\noindent{\bf 7. A LOWER BOUND ON THE BINDING ENERGY OF }
\noindent{\bf A SPHERICALLY SYMMETRIC DISTRIBUTION OF MATTER}
\vskip1pc

\vskip1pc
\noindent{\bf 7.1 Momentarily Static Spherically Symmetric Shell}
\vskip1pc

An important special case of the two-density model is the shell
with\footnote * {In the notation of Sect.5,
the shell corresponds to $\rho_1=0$, and the limiting
value $\sigma=\rho_2(\ell_2-\ell_1)$.}

$$\rho(\ell)=\sigma\delta (\ell-\ell_0)\,.\eqno(7.1)$$
This model will be useful in two regards. Firstly, it
represents the configuration with least binding energy. Secondly,
when we generalize our analysis to configurations with non-vanishing
current, it provides a useful exactly solvable model in which we
can examine the behavior of the geometry near singular points [3].

Inside the shell, space is flat so that $R=\ell$. As a consequence the
material energy $M$ is
equal to its Newtonian value

$$M=4\pi\sigma \ell_0^2\,.\eqno(7.2)$$
By integrating Eq.(6.2) across $\ell=\ell_0$ we find that
$R^\prime$
suffers a discontinuity at the shell given by

$$\Delta R^\prime = -4\pi\sigma \ell_0\,.\eqno(7.3)$$
Outside $R^\prime(\ell_{0+}) =1+\Delta R^\prime$.
If $\sigma$ is positive $R^\prime(\ell_{0+})$ will be bounded above by one.
As we have seen $M$ does not see the discontinuity in $R^\prime$.
The ADM mass $m_\infty$, however, does.

$$m_\infty =
m(\ell_{0+})={\ell_0\over 2}\Big(1 -R^{\prime2}(\ell_{0+})\Big)\,.
\eqno(7.4)$$

A horizon must form at some point outside the shell
if $4\pi\sigma \ell_0\ge 1$.

\noindent If $4\pi\sigma \ell_0< 2$,
$R^\prime >-1$,
$m_\infty$ is positive and the geometry is regular.

\noindent When $4\pi\sigma \ell_0 = 2$, $R^\prime=-1$, $m_infty=0$
and the geometry is weakly singular.
At this value, $R^\prime(\ell_{0+}) =-1$, the ADM mass $m$
vanishes and the geometry outside is flat and closed as well.
The bag of gold consistes of two flat caps sewn together along the
shell.

\noindent If $4\pi\sigma \ell_0 >2$, $R^\prime(\ell_{0+}) <-1$,
$m_\infty$ is negativeand the geometry suffers from
a strong singularity in the same way as it does
in the two-density model discussed in Sect.5.

A useful expression for $m_\infty$ is obtained by
writing $\Delta R^\prime$ in terms of $M$ and $\ell_0$. We use
(7.2) to express (7.3) $\Delta R^\prime = -M/\ell_0$
so that Eq.(7.4) can be rewritten in the form

$$m= M-{M^2\over 2\ell_0}\,.\eqno(7.5)$$
The binding energy $M-m=M^2/ (2\ell_0)$
is the Newtonian value. The distribution of matter with the least binding
energy in Newtonian gravity is a shell.

\vskip1pc
\noindent{\bf 7.2 Lower Bound on the Binding Energy}
\vskip1pc

It was conjectured by Arnowitt, Deser and Misner
but only proven recently by BM\'OM that, in general, [15, 10]

$$M-m\ge {M^2\over 2\ell} \,.\eqno(7.6)$$
This provides a lower bound on the binding energy of a spherically
symmetric distribution of matter of fixed $M$ and $\ell_0$.
The inequality is sharp. The distribution of matter which saturates
this inequality is the shell (compare Eq.(3.17)).

Note that if
$\ell$ is decreased while $M$ is kept fixed, the binding energy increases
--- the more compact the material system the larger its binding energy.
BM\'OM's proof relied on the use of a conformally flat coordinate system.
We will reproduce the proof using the proper radius directly.
The trick, as demonstrated by BM\'OM is to mimic the calculation for
the shell. We rewrite Eq.(6.2.2):

$$R^\prime={1\over R}\Gamma\,-\,{M\over R}\,.\eqno(7.7)$$
We now substitute for $R^\prime$ in the expression (2.5) for $m$:

$$m={R\over 2}\Big(1-R^{\prime2}\Big)=
	    -{M^2\over 2R} +{R\over 2}(1-F^2)
	                     +F)M\,,\eqno(7.8)$$
where we introduce the dimensionless quantity
$F:={\Gamma \over R}$. We now express the
difference  as the sum of the lower bound we wish
to establish and a remainder, $Q$

$$M-m={M^2\over 2\ell} + Q\,,\eqno(7.9)$$
where

$$Q\equiv {M^2\over2}\Big({1\over R}-{1\over \ell}\Big)
+(1-F)M +{R\over 2}(F^2-1)\,.\eqno(7.10)$$
We will show that $Q$ is always positive.
The key to doing this is to note that if the geometry is non-singular, then

$$1\le F \le {\ell\over R}\,.\eqno(7.11)$$
The upper bound is simply Eq.(6.2.4).\footnote *
{We also have already got a weak lower bound on $F$
which does not rely on the positivity of $\rho$,
$ \ell/ (2R)\le F$.}
To obtain the appropriate lower bound, let us first examine
the difference between the proper radius and the circumferential radius

$$\ell-R =\int_0^\ell d\ell (1-R^\prime)\,.$$
We can place a lower bound on this difference using the inequality
$1+R^\prime\le 2$:

$$\ell-R\ge {1\over 2}\int_0^\ell  d\ell(1-(R^\prime)^2)\,.$$
This can be inverted to provide the lower bound

$${1\over 2}\int_0^\ell d\ell\,(R^\prime)^2 \ge  R-{\ell\over2}\,.$$
{}From this we can deduce that $F\ge 1$.
We therefore have established both an upper bound and a lower bound on $F$.

We can now demonstrate that $Q$ is always positive.
We note that both the first and third terms are positive.
The discriminant of $Q$ is given by
$(F-1)\Big(F+1-2\ell/R\Big)$ and is negative. The quadratic therefore
possesses no real root. Thus $Q$ is positive everywhere.

Note that Eq.(7.6) implies $m\ge0$ iff $M \le 2\ell$.
This reproduces the sufficiency condition, Eq.(6.2.8).

\vfill
\eject
\centerline{\bf 8. CONCLUSIONS}
\vskip1pc

In this paper, we have attempted to identify generic features of
asymptotically flat spherically symmetric solutions to the
Hamiltonian constraint when the spatial geometry is momentarily static.

Our focus  has been on the characterization of
the two strong field features of such geometries,
apparent horizons and singularities. The simple exactly solvable models
consisting of piecewise constant energy density profiles
provide a useful guideline for the choice of appropriate variables
to characterize these features. Sufficiency conditions describing their
existence can be cast as inequalities between $M$ and $\ell_0$.
Matching inequalities of necessity are better cast in terms of
$\rho_{\rm Max}$ and $\ell_0$. We found that the latter inequalities
could be improved using simple functional inequalities. Because these
inequalities are not sharp, it is likely that they can be improved
with a more judicious exploitation of these inequalities.
This work might provide clues towards the identification
of appropriate variables with respect to which
to formulate the necessary part of the hoop conjecture [11].

The techniques introduced in this paper prepare the
ground for the examination of the constraints when
$K_{ab}\ne 0$. In paper IIb, we will examine the analogous
problem in this more general case.

At the end of paper I, we introduced the optical scalar plane
as a representation of the phase space of the theory.
In this representation non-singular solutions of the constraints
appear as bounded closed trajectories each containing the point
$P\equiv (2,2)$. The moment of time symmetry solutions we have considered
in detail in this paper correspond to degenerate trajectories that
run along the diagonal, $\omega_+=\omega_-$ starting out at the
point $P$. If the solution possesses a trapped surface it will
cross the origin, $(0,0)$, corresponding to $R^\prime=0$ at least once. If it
is non-singular, it will make an even number of crossings before returning
to the point $P$. In particular, we saw that when the energy density
profile is monotonically decreasing, a configuration can
cross the origin an arbitrarily large even number
of times. If the solution is singular it will cross the
origin an odd number of times before ultimately proceeding
towards the point, $Q\equiv (-2,-2)$. At this point,
the QLM vanishes and $R^\prime=-1$. Only in situations of
zero measure occuring when the energy density is fine-tuned
in such a way that the QLM tends monotonically to zero
will the trajectory terminate at this point. In general,
once $Q$ is breached the trajectory cannot re-enter the bounded
interval along the diagonal between $Q$ and $P$. What must occur is that
it continues monotonically along the diagonal towards
unboundedly large negative values of $R^\prime$.

When $K_{ab}\ne0$, there are many more possibilities [3].
Solutions may possess a future or a past apparent horizon.
These horizons will no longer generally coincide
with the extremal surfaces of the three-geometry.
Neither does the existence of the former
necessarily imply the existence of the latter.

Whereas in the MSCs, the only approach to singularity is through
the point $Q$, there is now a wide range of possibilities.
In general, a non-vanishing extrinsic curvature can lead to
more severe singularities than those encountered at a moment of time
symmetry. In addition, the constraints no longer imply that the scalar
curvature is finite. The converse of the positive QLM theorem
is no longer true. Indeed, $m$ can be positive everywhere
and yet the geometry be singular. $m$ will, however, always be finite
if the sources are finite.

What is remarkable is that it is possible to generalize the necessary and
sufficient conditions examined in this paper with the appropriate
generalizations of $M$ and $\rho_{\rm Max}$ and an appropriate
gauge. Not surprisingly, the gauges that do work are
procisely the $\alpha$ - parametrized linear extrinsic
curvature gauges introduced in paper I, with $\alpha$ in the range
$0.5 < \alpha <\infty$ [1].

The bags of gold which occur behind
singularities are physically disconnected from the
exterior and, as such, of little more than curiousity
value in the present context, at least, at the level of the
classical theory. In cosmology, however,
the bag of gold can be interpreted as a closed universe.

If $\ell=0$ is the north pole,
$\ell=\ell_S$ is just the south pole of the closed universe.
This closed universe will generally not be regular.
In fact, the argument presented in Sect.2 suggests that
regular closed universes constitute a subset of
zero measure in the set of all closed universes.
This universe will be singularity free if and only if the QLM vanishing
at one pole also vanishes at the other pole. We can think of the value
$M(\ell_S)$ assumed by $M$ at the south pole as a measure of the
binding energy of this regular closed universe.

Using Eq.(2.6$^\prime$), the integrability
condition $m(\ell_S)=0$ on a regular bag of gold implies

$$\int_0^{\ell_S} d\ell\, R^3 \rho^\prime=0\,.\eqno(8.1)$$
A regular spherically symmetric closed universe is impossible
if $\rho$ is strictly monotonic. $\rho$ must possess a maximum or a minimum
away from the poles. We will examine the configuration space of
regular spherically symmetric cosmologies in a subsequent paper [16].
\vskip 0.2cm
\centerline {\bf Acknowledgements}
\vskip1pc
This work was partially supported by Forbairt
grant SC/94/225. We wish to thank Edward Malec, many of the ideas discussed
here originated in discussions with him. We also would like to thank an
anonymous referee for many helpful suggestions.

\vskip2pc

\noindent{\bf APPENDIX I: THE RICCI TENSOR FROM THE SCALAR CURVATURE}
\vskip1pc

It is always possible to construct the Ricci tensor once we
know the scalar curvature.
We will do this for a general spherically symmetric geometry
by appealing to a mini-superspace Hilbert variational principle:
Let

$$I=\int d^3x\sqrt{g} {\cal R}\,.\eqno(a.1)$$
Then

$${1\over\sqrt{g}}{\delta I\over \delta g_{ab}}=
{\cal R}^{ab}-{1\over 2} {\cal R} g^{ab}\,.\eqno(a.2)$$
To implement the variational argument we need to restore the metric
coefficient ${\cal L}^2$ (defined in I) so that the
spatial coordinate system is only defined up to
spherical symmetry. This is because we require as many
independent metric components as there are independent components of
${\cal R}_{ab}$ which is two in this case.
Up to a divergence which is irrelevant for our purposes
(here $^\prime$ refers to differentiation with respect to $r$),

$$I= 2\int d^3x \left[{R^{\prime2}\over {\cal L}} +{\cal L}\right]\,.
\eqno(a.3)$$
Thus

$$\eqalign{{\delta I\over \delta R}
=&-4 \Big({R^{\prime}\over {\cal L}}\Big)^\prime\cr
{\delta I\over \delta {\cal L}}=&-\left[
\Big({R^{\prime}\over {\cal L}}\Big)^2 -{\cal L}\right]\,.\cr}
\eqno(a.4)$$
We now note that

$${\delta I\over \delta g_{rr}}
={1\over 2 {\cal L}}\Bigg({\delta I\over \delta {\cal L}}\Bigg)\,;\quad
{\delta I\over \delta g_{\theta\theta}}
={1\over 4R}\Bigg({\delta I\over \delta R}\Bigg)\,.
\eqno(a.5)$$
Having taken all variations, we can safely set ${\cal L}=1$.
We use Eqs.(a.2), (a.4) and (a.5) to express the non-vanishing
components of the Ricci tensor in the form

$${\cal R}^{rr} = {1\over 2} {\cal R}
+{1\over R^2}(R^{\prime2}-1)\,;\quad
{\cal R}^{\theta\theta} = {1\over 2R^2} {\cal R} +
{R^{\prime\prime}\over R^3}\,.\eqno(a.6)$$
We now eliminate the second derivative in the second term in favor of
${\cal R}$ and lower derivatives of $R$.
Eq.(2.12a \& b) for the scalars ${\cal R}_{\cal L}$
and ${\cal R}_R$ are then given by the appropriate projections.

In any regular solution of the constraints
${\cal R}_R\ge {\cal R}/4$, and
${\cal R}_{\cal L} \le {\cal R}/2$ everywhere.

We note that there is no non-trivial
regular everywhere solution of the Hamiltonian constraint
in which either ${\cal R}_R$ or ${\cal R}_{\cal L}$ vanishes.
In the case of ${\cal R}_R=0$ this is obvious on account of
the above bound. In the latter case,
consistency with the constraint (at a moment of time symmetry)
then implies that

$$\rho R^2 ={1\over R}\int_0^\ell \rho R^2 R^\prime d\ell\,,$$
so that $\rho\sim R^{-3}$ which is singular.
\vskip2pc
\noindent{\bf APPENDIX II: PROOF OF THE BOUND (6.3.13)}
\vskip2pc

In this appendix, we prove the validity of the bound,
Eq.(6.3.13) in the text for all $a>-1$.
For simplicity, we will first provide a proof for $a=1$.
The generalization will then be clear.
We define

$$ F(R, {\ell_1}) = \int_0^{\ell_1} R^3 d\ell \Big/\int_0^{\ell_1} R
d\ell\,.\eqno(b.1)$$
We will prove that

$$F(R,\ell_1)\le {\ell_1^2\over 2}\,,\eqno(b.2)$$
is true, for functions that satisfy

$$ R(0) = 0, \ \ \ R'(\ell) \le 1\,\ \ \ R \ge
0\,.\eqno(b.3)$$
Let us calculate the first variation of $F$

$$\delta F = \int_0^{\ell_1}(3R^2 - F)\delta R d\ell \Big/
 (\int_0^{\ell_1} R d\ell)^2\,.\eqno(b.4)$$
We claim that the maximum is achieved by $R = \ell$.

In this variational
problem, we can only consider variations which
satisfy Eq.(b.3), in other words we start with a function
$R_0$ which satisfies Eq.(b.3) and look for a family of
functions $R_t$, such that $R_{t = 0} = R_0$ and that for
some range of $t$, $0 \rightarrow \Delta t$, $R(t)$ also
satisfies Eq.(b.3). The $\delta R$ in Eq.(b.4) must be the
first derivative of such a sequence $R_t$ with respect to
$t$, evaluated at $t = 0$.

In particular, the allowed variations around $R = \ell$ must
satisfy

$$\delta R' \le 0, \ \ \ \delta R(0) = 0\,,\eqno(b.5)$$
which forces $\delta R$ to be negative and
monotonically decreasing on the interval (0, L). Now let
us consider the integral in Eq.(b.4), with $R = \ell$ and $F =
{\ell_1}^2/2$ if we replace $\delta R$ with a negative
constant $-C$. It is clearly negative, equalling $-C{\ell_1}^2/2$.
Notice that $3R^2 - F$ is negative in the range $(0,
{\ell_1}/\sqrt{6})$ and positive on the rest. This means that,
since $\delta R$ must be negative, the integrand in
Eq.(b.4) is positive on the interval $(0, {\ell_1}/\sqrt{6})$ and
negative elsewhere.

Since $\delta R$ is monotonically decreasing if we
replace $\delta R$ in the integral by its value at $\ell =
{\ell_1}/\sqrt{6}$ we make the integral more positive. However,
it is still negative. Therefore the actual integral is
negative and all the allowed variations reduce $F$. Hence
$R = \ell$ is a local maximum.

This is not enough, however. We need to show that no other
maximum occurs. To do this we note that $F < R^2_{\max}$, where
$R_{\max}$ is the maximum achieved by $R$ on the interval
$(0, {\ell_1})$. This can be shown by extracting  $R^2_{\max}$ from
the $R^3$ integral. We immediately get that $(3R^2 - F)$
must be positive near the maximum of $R$. If we further
have that $R' < 1$ at $R_{\max}$ we can find a positive
variation $\delta R$, localized near the maximum of
$R$, which satisfies the $R' <1$ condition. Such a
variation will increase $F$. Thus, in
particular, $R$ cannot possess an interior maximum. The
maximum must be assumed at $\ell={\ell_1}$ and at this maximum, $R'=1$.

Thus the only case we have to consider is the case where
$R$ achieves its maximum at ${\ell_1}$ and $R' = 1$ there.
We need only to show that if $R_{\max}< {\ell_1}$, there exists
a variation increasing $F$.

Suppose that $R_{\max}<{\ell_1}$. We cannot
have $R' = 1$ on the whole interval, for this
would imply $R = \ell$.
Therefore we have a point $\ell_2$ where $R(\ell_2)' < 1$. An
allowed variation is then $\delta R = 0$ on the
interval $(0, \ell_2)$ and $\delta R = 1$ on $(\ell_2, {\ell_1})$.
For $\ell_2$ close to ${\ell_1}$, the integral in Eq.(b.4) can
be approximated

$$\eqalign{\int_{\ell_2}^{\ell_1}[3(\ell - {\ell_1} + R_{\max})^2 - F] d\ell
= &R_{\max}^3 - (\ell_2 - {\ell_1} + R_{\max})^3 - F({\ell_1} - \ell_2)\cr
= &({\ell_1} - \ell_2)[({\ell_1} - \ell_2)^2 -3R_{\max}({\ell_1} - \ell_2) +
3R^2_{\max} - F]\,.\cr}\eqno(b.6)$$
We know that we can choose $\ell_2$ such that ${\ell_1} - \ell_2 <
R_{\max}$ because of the positivity of $R$. But this, in
conjunction with $F < R^2_{\max}$, is sufficient to show
that the expression in Eq.(b.6) is positive. This
completes the proof.

The generization of this proof is straightforward.
We can prove that

$$ F_{a}(R, {\ell_1}) = {\int_0^{\ell_1} R^{2 + a} d\ell \over
\int_0^{\ell_1} R^{a}d\ell}\, \le {(1 + a)\over 3 + a}{\ell_1}^2 \,,
\eqno(b.7)$$
for any $a > -1$. The technique is as before, first
of all we prove that $R = \ell$ is a local maximum, then we
prove that if $R_{\max} < {\ell_1}$ there exists an allowed
variation which increases $F$.

We have,

$$\delta F_{a} = {\int_0^{\ell_1}([2 + a]R^{1 +
a} - a FR^{-1 + a})\delta R d\ell
\over
(\int_0^{\ell_1} R^{a} d\ell)^2}\,.\eqno(b.8)$$
If $a < 0$ we immediately see that increasing $R$
increases $F$. Therefore $R = \ell$ is the maximum. We can
now restrict our attention to the case where $a \ge
0$.

If we replace $\delta R$ in Eq.(b.8) with a negative constant
$-C$ we can do the integration and show that it is clearly negative,
equalling $-2C{\ell_1}^2/(3 + a)$.
It is clear that the allowed $\delta R$'s must
be negative and monotonically decreasing. The integrand in
Eq.(b.8) is positive in the range $(0, \ell_2)$ and negative in
$(\ell_2, {\ell_1})$, where $\ell_2$ is given by

$$\ell_2^2 = {a (1 + a) \over (2 + a)(3 +
a)}{\ell_1}^2 <{\ell_1}^2\,.$$
Therefore we can replace $\delta R$ by its value at
$\ell_2$ and make the integral more positive. However, since
the integral with a negative constant is negative, the
correct integral is even more negative. Therefore $R = \ell$
is a local maximum.

The rest of the proof is just as before. We know that
$F < R^2_{\max}$ so therefore the quantity $([2 + a]R^
{1 + a} - a FR^{-1 + a})$ is positive near
the maximum of $R$. Therefore a positive variation,
localized near the maximum, increases $F$. The only
situation we need to deal with is where the maximum of
$R$ occurs at $\ell = {\ell_1}$ and simultaneously $R' = 1$ there.
Let us find the point $\ell = l_2$ where $R'(\ell_2) < 1$ and
$R = \ell - {\ell_1} + R_{\max}$ in the interval $(\ell_2, {\ell_1})$. Now an
allowed variation is one which is zero on the interval
$(0, \ell_2)$ and a positive constant on the interval $(\ell_2,
{\ell_1})$. For such a variation we can do the integration in
Eq.(b.8) to give

$$\eqalign{\delta F_{a}(\int_0^{\ell_1} R^{a} d\ell)^2 = &
R_{\max}^{2 + a} - (\ell_2 + R_{\max} - {\ell_1})^{2 + a}
- F R_{\max}^{a} + F(\ell_2 + R_{\max} - {\ell_1})^{a}\cr
= &  R_{\max}^{a}[R_{\max}^2 - F]
-(\ell_2 + R_{\max} - {\ell_1})^{a}[(\ell_2 + R_{\max} - {\ell_1})^2 -
F]\,.\cr} \eqno(b.9)$$
The first term in on the second line Eq.(b.9) is positive and also the first
term is greater than the second term because $R_{\max} >
\ell_2 + R_{\max} - {\ell_1}$. Therefore the variation of $F$ is
positive.

\vfill
\eject
\centerline{\bf Figure Captions}
\vskip2pc
\noindent{\bf Fig.(4.1) $M(\ell_0)/\ell_0$ vs. $\ell_0$}
Whereas $M$ is monotonic in $\ell_0$ by construction,
the ratio $M/\ell_0$ is not --- even before the geometry becomes
singular it begins to oscillate. The ratio rises monotonically for small
values of $\ell_0$, regardless of the  formation of an apparent horizon
until a maximum value is reached when the
surface becomes a decreasing point of inflection
of $R(\ell)$. Contrary to what we
might have expected, the maximum does not obtain with the
appearance of a singularity. As the star is
made even larger, the ratio begins to decrease
until it re-assumes the value $3/4$ as the surface becomes
singular. Subsequently, the ratio oscillates with decreasing amplitude with
one complete oscillation between each occurance of a singularity.
As $\ell_0\to\infty$, the ratio tends to the constant value
$3/4$ independent of $\rho$.

\vskip2pc
\centerline {\bf REFERENCES}
\vskip1pc
\item{1.} J. Guven and N \'O Murchadha (gr-qc/9411009, 1994)
subsequently referred to as I.
\vskip1pc
\item{2.} E. Malec and N. \'O Murchadha, {\it Phys. Rev.} {\bf 50}
R6033 (1994)
\vskip1pc
\item{3.} J. Guven and N \'O Murchadha (preprint, 1995)
subsequently referred to as III.
\vskip1pc
\item{4.} For an explicit example,
see E. Farhi, A.H. Guth and J. Guven {\it Nuc Phys} {\bf B339} 417 (1990)
\vskip1pc
\item{5.} See, for example, W. Unruh {\it Phys Rev} {\bf D 14} 870 (1976)
where the Hamiltonian is written down explicitly.
\vskip1pc
\item{6.} A detailed list of
relevant references to the definition of
the quasi-local mass in spherically symmetric
general relativity is provided in Ref.[1].
\vskip1pc
\item{7.} B.W. DeWitt {\it Phys. Rev} {\bf 160} (1966) 1113
\vskip1pc
\item{8.} P. Bizo\'n, E. Malec and N. \'O Murchadha {\it Phys. Rev. Lett.}
{\bf 61}, 1147 (1988).
\vskip1pc
\item{9.} P. Bizo\'n, E. Malec and N. \'O Murchadha {\it Class Quantum Grav}
{\bf 6}, 961 (1989)
\vskip1pc
\item{10.} P. Bizo\'n, E. Malec and N. \'O Murchadha {\it Class Quantum Grav}
{\bf 7}, 1953(1990)
\vskip1pc
\item{11.} See, for example,
E. Malec {\it Acta. Phys. Pol.} {\bf B22} 829 (1991);
E. Flanagan {\it Phys. Rev.} {\bf D44} 2409 (1991)
\vskip1pc
\item{12.} J. A. Wheeler,  {\it Geometrodynamics and the Issue of the
Final State} in Relativity, Groups and Topology, ed. by B.S. DeWitt and
C.M. DeWitt (Gordon and Breach 1964)
\vskip1pc
\item{13.} K. Kuchar and M. Ryan {\it Phys. Rev.} {\bf D40} 3982 (1990)
\vskip1pc
\item{14.} R. Schoen and S.-T.Yau {\it Commun. Math. Phys.}
{\bf 90} 575 (1983)
\vskip1pc
\item{15.} R. Arnowitt, S. Deser and C. Misner, {\it Ann Phys} {\bf 33} 88
(1966)
\vskip1pc
\item{16.} J. Guven and N \'O Murchadha (unpublished, 1994)

\bye